\documentclass[journal,twoside,web]{ieeecolor}
\usepackage{generic}
\usepackage{float}
\usepackage{cite}
\usepackage{amsmath,amssymb,amsfonts}
\usepackage{bbm}
\usepackage{bm}

\DeclareMathOperator*{\sgn}{sgn}
\usepackage{url}
\usepackage{algorithmic}
\usepackage{mathtools}
\usepackage{graphicx}
\usepackage{textcomp}
\def\BibTeX{{\rm B\kern-.05em{\sc i\kern-.025em b}\kern-.08em
    T\kern-.1667em\lower.7ex\hbox{E}\kern-.125emX}}
\usepackage{marginnote,zref-savepos}
\newcounter{labelnote}
\makeatletter
\let\oldmarginnote\marginnote
\renewcommand*{\marginnote}[1]{%
 \begingroup\strut
  \stepcounter{labelnote}\zsaveposx {marginnote-\thelabelnote}
     \ifnum 0\zposx{marginnote-\thelabelnote}<1900000
      \reversemarginpar
      \oldmarginnote{\color{blue}#1}%
     \else
      \normalmarginpar
      \oldmarginnote{\color{blue}#1}%
     \fi
 \endgroup%
}
\makeatother

\begin{document}
\title{Risk-sensitive safety analysis using\\Conditional Value-at-Risk*}
\author{Margaret P. Chapman, Riccardo Bonalli, Kevin M. Smith,\\Insoon Yang, Marco Pavone, Claire J. Tomlin 
\thanks{Received September 2019, revised January 2021, July 2021, and October 2021, and accepted November 2021. This work was supported in part by the National Science Foundation (NSF) under Grant NSF PIRE UNIV59732, by the National Cancer Institute (NCI) Cancer Systems Biology Consortium ``Measuring, Modeling and Controlling Heterogeneity'' through Oregon Health and Science University NCI \#1U54CA209988-01A1, and by the Computational Hydraulics International University Grant Program for complementary use of PCSWMM Professional software. M. P. Chapman was supported by the NSF Graduate Research Fellowship Program. K. M. Smith was supported in part by US National Science Foundation grant NSF-NRT 2021874. I. Yang is supported in part by the National Research Foundation of Korea funded by MSIT (2020R1C1C1009766).}
\thanks{M. P. Chapman is with the Edward S. Rogers Sr. Department of Electrical and Computer Engineering, University of Toronto, Toronto, Ontario M5S 3G4 Canada (email: mchapman@ece.utoronto.ca).} \thanks{R. Bonalli is with the Laboratory of Signals and Systems (L2S), Universit\'{e} Paris-Saclay, Centre National de la Recherche Scientifique (CNRS), CentraleSup\'{e}lec, France (email: riccardo.bonalli@centralesupelec.fr).} \thanks{K. M. Smith is with the Department of Civil and Environmental Engineering, Tufts University, Medford, MA 02155 USA and OptiRTC, Inc., Boston, MA 02116 USA (email:
kevin.smith@tufts.edu).} \thanks{I. Yang is with the Department of Electrical and Computer Engineering, and Automation and Systems Research Institute, Seoul National University, Seoul, South Korea (email: insoonyang@snu.ac.kr).} \thanks{M. Pavone is with the Department of Aeronautics and Astronautics, Stanford University, Stanford, CA 94305 USA (email: pavone@stanford.edu).} \thanks{C. J. Tomlin is with the Department of Electrical Engineering and Computer Sciences, University of California Berkeley, Berkeley, CA 94720 USA (email: tomlin@eecs.berkeley.edu).}
\thanks{*This paper was presented in part at the 2019 American Control Conference \cite{chapmanACC}.} 
}
\maketitle
\pagestyle{empty}
\thispagestyle{empty}

\begin{abstract}
This paper develops a safety analysis method for stochastic systems that is sensitive to the possibility and severity of rare harmful outcomes. We define \textit{risk-sensitive safe sets} as sub-level sets of the solution to a non-standard optimal control problem, where a random maximum cost is assessed via Conditional Value-at-Risk (CVaR). The objective function represents the maximum extent of constraint violation of the state trajectory, averaged over a given percentage of worst cases. This problem is well-motivated but difficult to solve tractably because the temporal decomposition for CVaR is history-dependent. Our primary theoretical contribution is to derive computationally tractable under-approximations to risk-sensitive safe sets. Our method provides a novel, theoretically guaranteed, parameter-dependent upper bound to the CVaR of a maximum cost without the need to augment the state space. For a fixed parameter value, the solution to only one Markov decision process problem is required to obtain the under-approximations for any family of risk-sensitivity levels. In addition, we propose a second definition for risk-sensitive safe sets and provide a tractable method for their estimation without using a parameter-dependent upper bound. The second definition is expressed in terms of a new coherent risk functional, which is inspired by CVaR. We demonstrate our primary theoretical contribution via numerical examples.
\end{abstract}
\newtheorem{algorithm}{Algorithm}
\newtheorem{theorem}{Theorem}
\newtheorem{remark}{Remark}
\newtheorem{lemma}{Lemma}
\newtheorem{assumption}{Assumption}
\newtheorem{definition}{Definition}
\begin{figure}[h]
\centerline{\includegraphics[width=0.8\columnwidth]{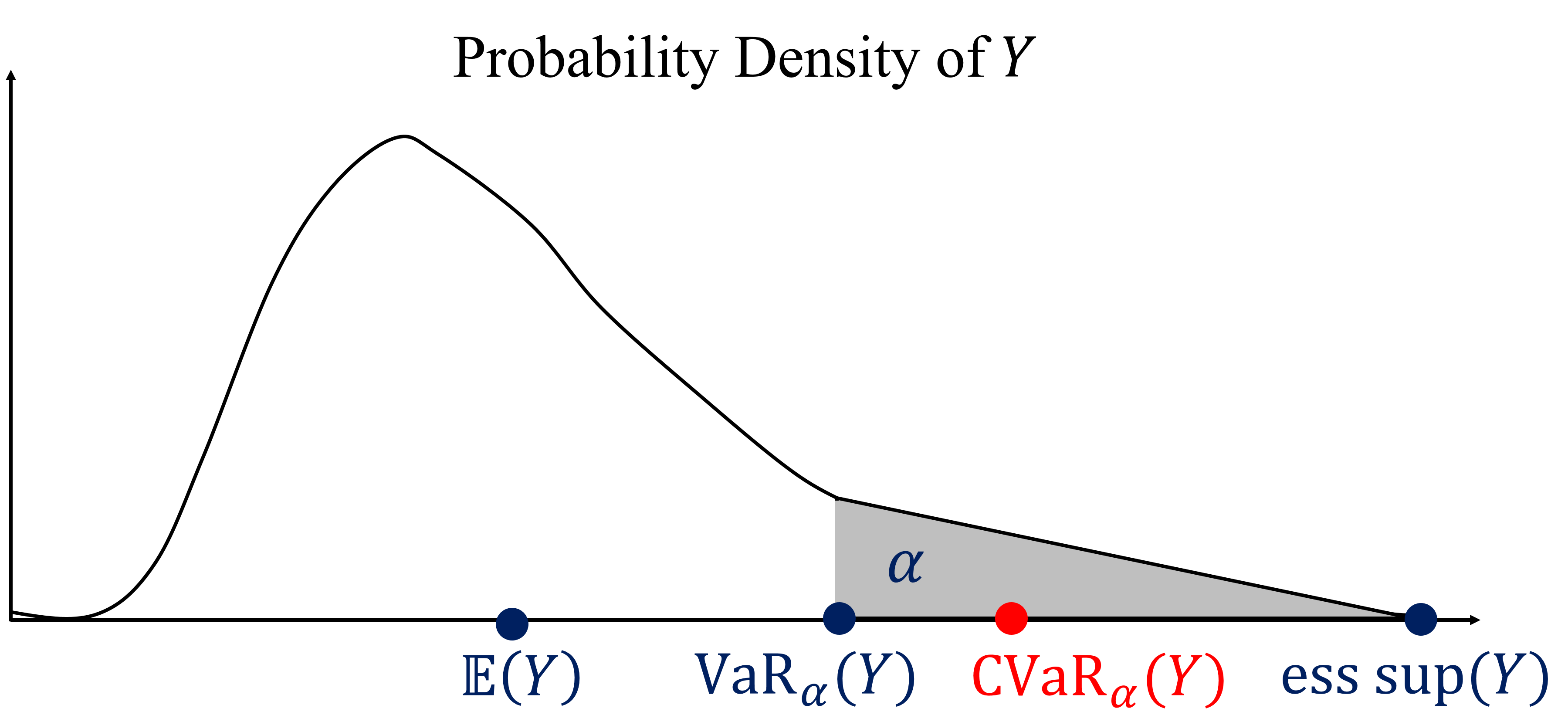}}
\caption{The Conditional Value-at-Risk (CVaR) quantifies the upper tail of a cost distribution. For an absolutely continuous, bounded random variable $Y$ representing a cost and $\alpha \in (0,1]$, we illustrate the expected cost in the $\alpha\cdot 100$\% worst cases, which is $\text{CVaR}_\alpha(Y)$ in this setting. The area of the shaded region is $\alpha$. The expectation of $Y$, the Value-at-Risk of $Y$ at level $\alpha$ (the lowest cost in the $\alpha \cdot 100$\% worst cases), and the essential supremum of $Y$ are also shown.}
\label{cvarpic}
\end{figure}
\begin{figure}[h]
\centerline{\includegraphics[width=0.7\columnwidth]{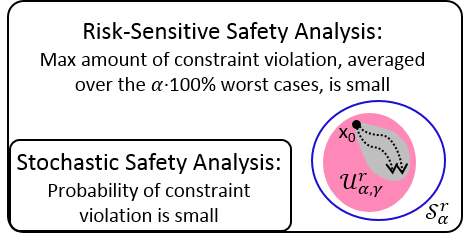}}
\caption{We develop a safety analysis method that generalizes stochastic safety analysis by assessing \textcolor{black}{the severity of random} harmful outcomes. We define the risk-sensitive safe set $\mathcal{S}_\alpha^r$ in terms of CVaR and derive an under-approximation $\mathcal{U}_{\alpha,\gamma}^r$ that is computationally tractable. $\mathcal{S}_\alpha^r$ represents the set of initial states from which the maximum extent of constraint violation of the state trajectory, averaged over the $\alpha \cdot 100$\% worst cases, can be reduced to a threshold $r$.}
\label{intpic}
\end{figure}
\section{Introduction}
\IEEEPARstart{C}{ontrol}-theoretic formal verification methods for dynamical systems typically fall in the robust domain~\cite{bertsekas1971minimax, mitchell2005time, lygeros2011, chen2018decomp, chen2018hamilton} or in the stochastic domain~\cite{abate2008probabilistic,summers2010verification, kambarpour2011, ding2013}. Robust methods for formal verification assume that uncertain disturbances lack probabilistic descriptions, live in bounded sets, and exhibit adversarial behavior. These assumptions are appropriate if probabilistic information about disturbances is not available, and if the conservative policy or safety specification that results from a pessimistic world view is useful in practice. However, when one considers formal verification as a design tool for safety-critical systems in the digital world today, it is reasonable to assume that simulation tools or sensor data are available to estimate probabilistic descriptions for disturbances. Moreover, it is reasonable to consider the following world view: disturbances need not be adversarial, but rare harmful outcomes are still possible. 

Control-theoretic stochastic formal verification methods do assume that disturbances are probabilistic and can be non-adversarial~\cite{abate2008probabilistic,summers2010verification} or adversarial~\cite{kambarpour2011, ding2013} in nature. These methods compute the probability of safety or performance by using expected indicator cost functions. The expectation, however, is not designed to quantify the features in the tails of a distribution, and the probability of a harmful outcome need not indicate its severity. Thus, formal verification methods at the intersection of the robust and stochastic domains are emerging. A method for distributionally robust safety analysis has been proposed \cite{yang2018dynamic}, and methods that use risk measures to assess harmful tail costs, e.g., \cite{samuelson2018safety} and our prior work \cite{chapmanACC}, have been introduced.\footnote{A \emph{risk measure} (risk functional) is a map from a set of random variables to the extended real line. Exponential utility, Value-at-Risk, CVaR, and Mean-Deviation are examples \cite{shapiro2009lectures}. The terms risk measure and risk functional are interchangeable.}

While the notion of risk-sensitive formal verification is recent, it is related to the notion of risk-sensitive Markov decision processes (MDPs), which dates back to the early 1970s. In 1972, Howard and Matheson studied risk-sensitive MDPs on finite state spaces, where the cost is evaluated in terms of exponential utility \cite{howardmat1972}. This idea was transferred to \textcolor{black}{linear} control systems by Jacobson in 1973 \cite{jacobson1973} and was further developed in later decades. For example, see the seminal works by Whittle \cite{whittle1981, whittle1990risk} and di Masi and Stettner \cite{masi1999}. \textcolor{black}{The exponential utility of a non-negative random cost $Y$ $\mathcal{J}_\theta(Y) \coloneqq \frac{-2}{\theta}\log(E(e^{\frac{-\theta}{2}Y}))$ assesses the risk of $Y$ in terms of the moments of $Y$ and is parametrized by a non-zero scalar $\theta$. Under appropriate conditions, $\mathcal{J}_\theta(Y)$ tends to $E(Y)$ as $\theta \rightarrow 0$ and $\mathcal{J}_\theta(Y) \approx E(Y) - \frac{\theta}{4}\text{Variance}(Y)$ if $|\theta|$ is sufficiently small \cite{whittle1981}. The risk-averse setting corresponds to $\theta < 0$. 
However, if $\theta$ is too negative, the controller can suffer from a phenomenon called ``neurotic breakdown'' in the linear-quadratic-Gaussian setting \cite{whittle1981}.}

Hence, the notion of risk-sensitive MDPs has been generalized beyond exponential utility. Kreps used the expectation of a utility function as a risk-sensitive performance criterion for MDPs \cite{Kreps1977}. Ruszczy\'{n}ski defined a risk-sensitive performance criterion for MDPs in terms of a composition of risk measures \cite{ruszczynski2010risk}. State-space augmentation has been used to optimize the cumulative cost of a MDP, where the cost is assessed via CVaR \cite{bauerleott} or a certainty equivalent risk measure \cite{bauerlerieder}. The former problem is called a CVaR-MDP. Convex analytic methods have been used to solve MDPs with expected utility or CVaR criteria via state-space augmentation and infinite-dimensional linear programming \cite{haskell}. A temporal decomposition for CVaR \cite{pflug2016timeEuro, pflug2016timeMath} has been used to propose a dynamic programming (DP) algorithm on an augmented state space to solve a CVaR-MDP problem approximately \cite{chow2015risk}. \textcolor{black}{Analysis at the intersection of mean field games, linear systems, and risk measures with connections to CVaR is provided by \cite{meanfield}.}

\textcolor{black}{Ruszczy\'{n}ski's approach \cite{ruszczynski2010risk} and MDPs that assess cumulative costs via expectation or exponential utility are \emph{time-consistent} problems. That is, these problems satisfy Bellman's Principle of Optimality on the original state space.\footnote{Different meanings for time consistency have been proposed, e.g., see \cite{ruszczynski2010risk, shapiro2009timeconsistency, bodafilar2006}. We refer to the meaning for time consistency from \cite{bodafilar2006}.} However, a CVaR-MDP is time-inconsistent. Several solution concepts for time-inconsistent problems have been proposed. For example, a game-theoretic solution concept is studied in \cite{bjork}, which considers the problem as a game against one’s future self. Another popular approach is to focus on \emph{pre-commitment strategies} that cannot be revised at later stages. Optimal \textcolor{black}{or nearly optimal} pre-commitment strategies can be obtained using the structure of CVaR; see \cite{bauerleott, haskell, yang2017}, for example. Although an optimal pre-commitment strategy is globally optimal only at the initial stage, maintaining suitable empirical performance at later stages is possible, particularly when the time horizon is not too long \cite{Rudloff2014}. In mean-CVaR asset allocation problems, optimal pre-commitment strategies are shown to be effective even with long time horizons \cite{Forsyth2020}.}

A line of research that falls between risk-sensitive MDPs and standard risk-neutral MDPs is risk-constrained MDPs \cite{borkar, van2015distributionally, samuelson2018safety, haskell}. Here, the goal is to minimize an expected cumulative cost subject to a risk constraint that limits the extent of a cost. Refs. \cite{borkar, van2015distributionally, samuelson2018safety}, for example, express this constraint in terms of CVaR. 

The additional effort required to solve time-inconsistent problems, including CVaR-MDPs, may be justified for safety-critical applications. A strong theoretical basis for using CVaR to assess harmful tail costs has been in development since the early 2000s, e.g., see \cite{rockafellar2002conditional} and the references therein. \textcolor{black}{Informally}, CVaR represents the expected cost in the $\alpha\cdot 100$\% worst cases, where $\alpha \in (0,1]$ (Fig. \ref{cvarpic}). CVaR quantifies the more harmful tail of a distribution, and managing this tail is paramount in safety-critical applications. 
%

This paper proposes a method to assess how well a stochastic system can remain within a desired operating region with respect to a range of worst-case perspectives. We call this method \emph{risk-sensitive safety analysis} (Fig. \ref{intpic}). Its foundation is a non-standard optimal control problem that evaluates a random maximum cost via CVaR. The objective function represents the maximum extent of constraint violation of the state trajectory, averaged over the $\alpha\cdot 100$\% worst cases, where $\alpha \in (0,1]$ is a \emph{risk-sensitivity level}. This problem is difficult to solve tractably because the temporal decomposition for CVaR is history-dependent \cite{pflug2016timeEuro, pflug2016timeMath}. We define \textit{risk-sensitive safe sets} as sub-level sets of the solution to this non-standard problem. These sets are powerful tools for safety analysis. Indeed, they assess system behavior on a spectrum of worst cases, while being sensitive to the possibility and severity of rare harmful outcomes. 

Our primary theoretical contribution is to derive computationally tractable under-approximations to risk-sensitive safe sets. We derive these under-approximations by proving the following: for any control policy and any initial state, the CVaR of a maximum cost is upper bounded by a scaled logarithm of an expected cumulative cost, where the stage cost has a specific analytical form. For this proof, we use various properties of CVaR and the log-sum-exponential approximation to the maximum. The latter approximation depends on a parameter, $\gamma \in \mathbb{R}$. For a fixed $\gamma$, the solution to one MDP problem is required to obtain the under-approximations for any family of risk-sensitivity levels. We provide practical insights on how to choose such a parameter in the experimental section.

Our method provides a novel, theoretically guaranteed upper bound to the CVaR of a \emph{maximum} cost for the purpose of \emph{safety analysis} without the need to augment the state space. (Augmenting the state space may be less tractable in some settings, e.g., when the range of the augmented state is large.) In contrast, existing methods aim to compute the CVaR of a cumulative cost via state-space augmentation. By taking different approaches to augment the state space, Refs. \cite{bauerleott} and \cite{haskell} minimize the CVaR of a cumulative cost, and Ref. \cite{chow2015risk} minimizes the CVaR of a cumulative cost approximately. These related works are focused on controller synthesis but are not focused on safety analysis.

Our secondary theoretical contribution is to propose a second definition for risk-sensitive safe sets and provide a tractable method for their estimation without using a parameter-dependent upper bound. The second definition is expressed in terms of a new risk functional, which is inspired by CVaR and has certain desirable properties. In particular, we prove that this risk functional admits an upper bound that can be computed via DP (on the original state space and without an additional parameter that requires tuning). This result forges a new path to estimate risk-sensitive safety criteria with desirable computational attributes.

\emph{Organization}. We present notation and background on CVaR in Sec. \ref{cvarback}. Our primary and secondary theoretical contributions are provided in Sec. \ref{theory} and Sec. \ref{sectheory}, respectively. We develop computational examples of a temperature system and a stormwater system to demonstrate our primary theoretical contribution in Sec. \ref{num}. Sec. \ref{conclusion} presents conclusions and directions for future work. 
%
\section{Background on Conditional Value-at-Risk}\label{cvarback}
We use the following notation. If $S$ is a metrizable space, $\mathcal{B}(S)$ is the Borel sigma algebra on $S$.  If $(\Omega, \mathcal{F}, \mu)$ is a probability space and $1 \leq p \leq \infty$, $L^p(\Omega,\mathcal{F},\mu)$ is the associated $L^p$ space, and $||\cdot||_p$ is the associated norm. 
Typically, we use upper-case letters to denote random variables or sets, whereas lower-case letters denote deterministic quantities, including parameters. Exceptions are the length of a time horizon is expressed in terms of $T \in \mathbb{N}$ and $E(\cdot)$ denotes expectation. 

Next, we present a standard definition for CVaR and facts about CVaR that are relevant to this work.\footnote{Additional names for CVaR include Average Value-at-Risk, Expected Shortfall, and Expected Tail Loss. We present the definition for CVaR that is used by Shapiro and colleagues, e.g., \cite{shapiro2009lectures, xinshapiro2012,shapiro2012}.} Let $Y$ be a random variable with finite first moment, representing a cost, defined on a probability space $(\Omega, \mathcal{F}, \mu)$. That is, let $Y \in L^1(\Omega,\mathcal{F},\mu)$, where smaller realizations of $Y$ are preferred. The \emph{Conditional Value-at-Risk} of $Y \in L^1(\Omega,\mathcal{F},\mu)$ at the \emph{risk-sensitivity level} $\alpha \in (0,1]$ is defined by
\begin{equation}\label{cvardef}
    \text{CVaR}_{\alpha}(Y) \coloneqq \inf_{s \in \mathbb{R}} \Big( s + \textstyle\frac{1}{\alpha} E(\max(Y-s,0)) \Big),
\end{equation}
where $E(\cdot)$ is the expectation with respect to (w.r.t.) $\mu$. 
We note the following consequences of the Definition \eqref{cvardef}:
\begin{enumerate}
    \item $\text{CVaR}_{1}(Y) = E(Y)$.
    \item If $0 < \alpha_1 \leq \alpha_2 \leq 1$, then $\text{CVaR}_{\alpha_1}(Y) \geq \text{CVaR}_{\alpha_2}(Y)$ and $\text{CVaR}_{\alpha_i}(Y) \in \mathbb{R}$ for $i=1,2$.
\end{enumerate}
Definition \eqref{cvardef} is not the most intuitive, so we present an alternative definition that explains the names Conditional Value-at-Risk and Average Value-at-Risk. The alternative definition is written in terms of the Value-at-Risk of $Y \in L^1(\Omega,\mathcal{F},\mu)$ at level $\alpha \in (0,1)$, which is given by
\begin{equation}
    \text{VaR}_{\alpha}(Y) \coloneqq \inf \big\{y \in \mathbb{R} : \mu(\{Y\leq y\}) \geq 1-\alpha \big\},
\end{equation}
where $\mu(\{Y\leq y\})$ is the probability of the event $\{Y\leq y\} \coloneqq \{ \omega \in \Omega : Y(\omega)\leq y\} \in \mathcal{F}$. In other words, $\text{VaR}_{\alpha}(Y)$ is the generalized inverse cumulative distribution function of $Y$ at level $1-\alpha$, or equivalently, the left-side $(1-\alpha)$-quantile of the distribution of $Y$ \cite{shapiro2012}. 
The CVaR of $Y \in L^1(\Omega,\mathcal{F},\mu)$ at level $\alpha \in (0,1)$ is equivalent to an average of the Value-at-Risk \cite[Thm. 6.2]{shapiro2009lectures}:
\begin{equation}\label{2b}
    \text{CVaR}_{\alpha}(Y) = 
      \frac{1}{\alpha} \int_{1-\alpha}^1 \text{VaR}_{1-p}(Y) \; \mathrm{d}p.
    \end{equation}
The above equation explains the commonly used name \emph{Average Value-at-Risk}. Now, to explain the name \emph{Conditional Value-at-Risk}, suppose that the cumulative distribution function $F_Y(y) \coloneqq \mu(\{Y\leq y\})$ is continuous at $y = \text{VaR}_\alpha(Y)$. Continue to assume that $Y \in L^1(\Omega,\mathcal{F},\mu)$ and $\alpha \in (0,1)$. Then, $\text{CVaR}_{\alpha}(Y)$ is a \emph{conditional} expectation that is expressed in terms of the \emph{Value-at-Risk} \cite[Thm. 6.2]{shapiro2009lectures}:
\begin{equation}\label{nameforcvar}
    \text{CVaR}_{\alpha}(Y) = E( Y \hspace{-.6mm} \mid \hspace{-.5mm} Y \geq \text{VaR}_\alpha(Y) ).
\end{equation}
Equation \eqref{nameforcvar} means that $\text{CVaR}_{\alpha}(Y)$ represents the expected value of $Y$ in the $\alpha \cdot 100 \%$ worst cases.

CVaR is a commonly cited example of a \emph{coherent} risk functional \cite{shapiro2009lectures, shapiro2012}. Coherent risk functionals are a class of risk functionals, first proposed by Artzner et al. \cite{Artzner}, that satisfy four properties, which are particularly meaningful in applications where sensitivity to risk is critical. We present these properties in the context of CVaR at level $\alpha \in (0,1]$, where $Y_i \in L^1(\Omega,\mathcal{F},\mu)$ below.
\begin{enumerate}
    \item \emph{Monotonicity}. If $Y_1(\omega) \leq Y_2(\omega)$ for almost every (a.e.) $\omega \in \Omega$, then $\text{CVaR}_{\alpha}(Y_1) \leq \text{CVaR}_\alpha(Y_2)$. That is, a random cost that is larger than another almost everywhere incurs a larger risk.
    \item \emph{Subadditivity}. $\text{CVaR}_{\alpha}(Y_1 + Y_2) \leq \text{CVaR}_{\alpha}(Y_1) + \text{CVaR}_{\alpha}(Y_2)$. If $Y_i$ is the (random) stage cost of a control system at time $i$, then the risk of the cumulative cost over a finite horizon is at most the \textcolor{black}{sum of the risks of the stage costs}.
    \item \emph{Translation equivariance}. If $a \in \mathbb{R}$, then $\text{CVaR}_{\alpha}(Y_1 + a) = \text{CVaR}_{\alpha}(Y_1) + a$.
    \item \emph{Positive homogeneity}. If $0\leq \lambda < \infty$, then $\text{CVaR}_{\alpha}(\lambda Y_1) = \lambda\text{CVaR}_{\alpha} (Y_1)$.
\end{enumerate}
The last two properties ensure that shifting or scaling a random variable provides an analogous transformation to the risk of the random variable. In particular, the expectation operator satisfies the four properties above and thus is a coherent risk functional. We use some of these properties in our proofs. We also use the fact that a real-valued coherent risk functional can be represented in terms of a supremum over a family of expectations.\footnote{The family of expectations has specific properties that are out of the scope of this paper. The representation was developed over several years, e.g., see \cite{Artzner, follmer2004, rushapiro2006, shapiro2012}.} This representation takes the following form for CVaR at level $\alpha \in (0,1]$ \cite{shapiro2012}: for any $Y \in L^1(\Omega,\mathcal{F},\mu)$,
\begin{subequations}\label{12}
\begin{equation}
    \text{CVaR}_\alpha(Y) = \sup_{Q \in \mathcal{Q}_\alpha} \;  \int_\Omega  Y \; \mathrm{d}Q = \sup_{\xi \in \mathcal{A}_\alpha}   \int_\Omega Y  \xi \; \mathrm{d}\mu ,
\end{equation}
where the definitions of $\mathcal{Q}_\alpha$ and $\mathcal{A}_\alpha$ follow. $Q \in \mathcal{Q}_\alpha$ if and only if $Q$ is a probability measure that is absolutely continuous with respect to $\mu$, i.e., of the form
    $\textstyle Q(B) = \int_B \xi \mathrm{d}\mu$, where $B \in \mathcal{F}$ and
$\xi \in \mathcal{A}_\alpha$.
$\mathcal{A}_\alpha$ is a set of densities defined by
\begin{equation}\label{12c}
    \mathcal{A}_\alpha \coloneqq \left\{\xi \in L^\infty(\Omega,\mathcal{F},\mu) : 0 \leq \xi \leq \frac{1}{\alpha} \text{ a.e.}, \; \int_\Omega \xi \; \mathrm{d}\mu = 1 \right\}.
\end{equation}
\end{subequations}
%
\section{CVaR-Based Risk-Sensitive Safety Analysis}\label{theory}
We use the CVaR functional to pose a safety analysis problem. We consider a stochastic system evolving on a discrete, finite-time horizon and start with the standard set-up for this setting. Let $S$ and $A$ be Borel spaces, representing the set of states and the set of controls of the system, respectively.
Define the sample space $\Omega \coloneqq (S \times A)^T \times S$, where $\omega \coloneqq (x_0,u_0,\dots,x_{T-1},u_{T-1},x_T) \in \Omega$ is a finite sequence of states and controls that may be realized on a time horizon of length $T+1$ and $T \in \mathbb{N}$ is given. The random state $X_t : \Omega \rightarrow S$ and the random control $U_t : \Omega \rightarrow A$ are projections. That is, for any $\omega \in \Omega$ of the form above, define $X_t(\omega) \coloneqq x_t$ and $U_t(\omega) \coloneqq u_t$, where the coordinates of $\omega$ have casual dependencies, to be described. The initial state $X_0$ is fixed arbitrarily at $x \in S$. 
%
The system's evolution is affected by $W$-valued random disturbances $(D_0,D_1,\dots,D_{T-1})$ with a common distribution $P_D$, where $W$ is a Borel space. $D_t$ is independent of the states, controls, and $D_s$ for any $s \neq t$.
The distribution of $X_{t+1}$ conditioned on $(X_t,U_t) = (x_t,u_t) \in S \times A$ is defined as follows: for any $B \in \mathcal{B}(S)$,
\begin{equation}\label{mytransitionlaw}
Q(B | x_t,u_t) \coloneqq P_D\big(\{ d_t \in W : f(x_t,u_t,d_t) \in B \}\big),
\end{equation} 
where $f : S \times A \times W \rightarrow S$ is a Borel-measurable map that models the system dynamics. 
%
We use the typical class of random, history-dependent policies $\Pi$. Each $\pi \in \Pi$ takes the form
    $\pi = (\pi_0,\pi_1,\dots,\pi_{T-1})$,
where each $\pi_t$ is a Borel-measurable stochastic kernel on $A$ given $H^t \coloneqq (S \times A)^t \times S$. 

The above set-up is standard in discrete-time stochastic control. One reason is that, given $x \in S$ and $\pi \in \Pi$, the set-up allows the construction of a unique probability measure $P_x^\pi$ that characterizes the system's evolution, provided that the system is initialized at $x$ and uses the policy $\pi$ \textcolor{black}{(Ionescu-Tulcea Theorem)}. The measure $P_x^\pi$ permits the prediction of the system's performance over time under uncertainty. 
Random costs incurred by the system are defined on $(\Omega,\mathcal{B}(\Omega), P_x^\pi)$, a probability space parametrized by $x$ and $\pi$.
The notation $E_x^\pi(\cdot)$ is the expectation operator with respect to $P_x^\pi$.
\subsection{On Evaluating a Random Cost via CVaR}
We use $(\Omega,\mathcal{B}(\Omega), P_x^\pi)$ to define a random cost for the system and to evaluate this cost via CVaR. Suppose that there is a \emph{constraint set} $K \in \mathcal{B}(S)$, where the state trajectory $(X_0,X_1,\dots,X_{T})$ of the system should remain inside. It may be impossible for the system to remain inside $K$ always due to random disturbances in the environment. Let $g_K: S \rightarrow \mathbb{R}$ be a bounded Borel-measurable function that represents a notion of distance between a state realization and the boundary of $K$. Specifically, $g_K(x_t)$ is the \emph{extent of constraint violation} of $x_t$, a realization of the random state $X_t$. More specifically, if $x_t$ is outside of $K$ and far from the boundary of $K$, then $g_K(x_t)$ has a large positive value. However, if $x_t$ is inside of $K$, then $g_K(x_t)$ may be 
\begin{enumerate}
    \item zero, if one does not favor certain trajectories inside of $K$, or
    \item a more negative value when $x_t$ is more deeply inside of $K$, if one favors trajectories that remain deeply inside of $K$.
\end{enumerate}
Using $g_K$, we define a random $\mathbb{R}$-valued cost that quantifies the \emph{maximum extent of constraint violation of the state trajectory}: for any $\omega = (x_0,u_0,\dots,x_{T-1},u_{T-1},x_T) \in \Omega$,
\begin{equation}\label{G}
    G(\omega) \coloneqq \max_{t = 0,1,\dots,T} \; g_K(X_t(\omega)) = \max_{t = 0,1,\dots,T} \; g_K(x_t).
\end{equation}
In other words, $G$ quantifies how well the random state trajectory satisfies the safety criterion to remain inside of $K$. Hence, $G$ quantifies the safety of the random state trajectory, which is defined with respect to the constraint set $K$ via the function $g_K$. A deterministic (and continuous-time) version of \eqref{G} is used in Hamilton-Jacobi reachability analysis, a robust safety analysis method for (non-stochastic) uncertain systems, which has been established over the past 15 years; e.g., see \cite{mitchell2005time, kenethesis, chen2018hamilton}, and the references therein. A standard choice for $g_K$ is a clipped signed distance function with respect to $K$ \cite[p. 8]{kenethesis}. In our numerical example of a thermostatically controlled load, we use $g_K(x_t) = \max(x_t - 21, 20 - x_t)$ to quantify how far a state realization $x_t$ can be inside or outside of $K = [20, 21]$ $^\circ$C (Sec. \ref{tempsysex}).
%

It holds that $G \in L^\infty \coloneqq L^\infty(\Omega,\mathcal{B}(\Omega),P_x^\pi)$ and $G \in L^1 \coloneqq L^1(\Omega,\mathcal{B}(\Omega),P_x^\pi)$. The function $g_K$ composed with $X_t$ is an element of $L^\infty$ because $g_K : S \rightarrow \mathbb{R}$ is bounded and Borel measurable and $X_t : \Omega \rightarrow S$ is Borel measurable. Thus, $G$ is a point-wise maximum of finitely many functions in $L^\infty$. Therefore, $G$ inherits the measurability properties of these functions and is essentially bounded. Since $(\Omega,\mathcal{B}(\Omega),P_x^\pi)$ is a probability space, $L^\infty$ is a subset of $L^1$, and it follows that $G \in L^1$ as well. 

Now, 
we express the CVaR of $G$. The CVaR of $G \in L^1(\Omega,\mathcal{B}(\Omega),P_x^\pi)$ at level $\alpha \in (0,1]$ is given by
\begin{subequations}\label{cvardefwithG}
\begin{equation}
    \text{CVaR}_{\alpha,x}^\pi(G) \coloneqq \inf_{s \in \mathbb{R}} \Big( s + \textstyle\frac{1}{\alpha} E_x^\pi(\max(G-s,0)) \Big).
\end{equation}
By using \eqref{12}, it holds that
\begin{equation}\label{9b}
    \text{CVaR}_{\alpha,x}^\pi(G) = \sup_{\xi \in \mathcal{A}_{\alpha,x}^\pi}  \int_\Omega G  \xi \; \mathrm{d}P_x^\pi,
\end{equation}
where $\mathcal{A}_{\alpha,x}^\pi$ is a set of densities defined by
\begin{equation}\label{9c}
    \mathcal{A}_{\alpha,x}^\pi \hspace{-.9mm} \coloneqq \hspace{-.9mm} \big\{\xi \in L^\infty(\Omega,\mathcal{B}(\Omega),P_x^\pi) : 0 \leq \xi \leq \textstyle\frac{1}{\alpha} \text{ a.e.},  E_x^\pi(\xi) = 1 \big\}.
\end{equation}
\end{subequations}
We use \eqref{G} and \eqref{cvardefwithG} to define risk-sensitive safe sets next.
\subsection{Risk-Sensitive Safe Sets}
\begin{definition}[Risk-Sensitive Safe Sets]\label{risksafedef}
Let $\alpha \in (0,1]$ and $r \in \mathbb{R}$ be given. The \emph{$(\alpha,r)$-risk-sensitive safe set for a given policy $\pi \in \Pi$} is defined by
\begin{equation}\label{mySgivenpolicy}\begin{aligned}
\mathcal{S}_{\alpha}^{r,\pi} & \coloneqq \left\{x \in S : \text{CVaR}_{\alpha,x}^\pi\Big(\max_{t = 0,1,\dots,T}  g_K(X_t)\Big) \leq r \right\}.
\end{aligned}\end{equation}
The \emph{$(\alpha,r)$-risk-sensitive safe set} is defined by
\begin{equation}\label{myS}\begin{aligned}
\mathcal{S}_{\alpha}^{r} & \coloneqq \left\{x \in S : \inf_{\pi \in \Pi} \text{CVaR}_{\alpha,x}^\pi\Big(\max_{t = 0,1,\dots,T}  g_K(X_t)\Big) \leq r \right\}.
\end{aligned}\end{equation}
\end{definition}
We denote the infimum in \eqref{myS} by $W_\alpha^*(x)$. Risk-sensitive safe sets are well-motivated. These sets represent the sets of initial states from which the maximum extent of constraint violation of the state trajectory, averaged over the $\alpha \cdot 100$\% worst cases, can be made sufficiently small. The maximum extent of constraint violation of the state trajectory is the real-valued random variable $G \coloneqq \max_{t = 0,1,\dots,T} g_K(X_t)$. We allow $g_K$ to be negative so that decision-makers can encode preferences for trajectories remaining deeper inside of $K$ over trajectories near the boundary of $K$, if desired. In our numerical example of a thermostatically controlled load, we allow $g_K$ to take on both negative and non-negative values to express a preference for trajectories that remain closer to 20.5 $^{\circ}\text{C}$ (Sec. \ref{tempsysex}). In our numerical example of a stormwater system, however, we choose a non-negative $g_K$ to utilize all capacity in the water storage tanks without penalty (Sec. \ref{watersysex}).
%

Using CVaR to define risk-sensitive safe sets is well-justified from a decision-theoretic point of view because CVaR is a coherent risk measure. That is, CVaR satisfies the axioms of monotonicity, subadditivity, positive homogeneity, and translation equivariance. Sec. \ref{cvarback} provides intuitive interpretations for these axioms. Besides having an axiomatic justification, CVaR has the useful interpretation of quantifying the upper tail of a distribution. Indeed, CVaR provides a \emph{quantitative characterization of risk aversion} by representing the expected cost in the $\alpha \cdot 100$\% worst cases, where $\alpha \in (0,1]$ is selected by the decision-maker. This interpretation is exact if continuous random variables \textcolor{black}{in $L^1$} are evaluated.

Risk-sensitive safe sets generalize probabilistic safe sets \cite{abate2008probabilistic} by quantifying the maximal \emph{extent} of constraint violation at a given risk-sensitivity level rather than the probability of constraint violation. Risk-sensitive safe sets quantify \emph{how much} constraint violation occurs on average in the $\alpha \cdot 100\%$ worst cases, whereas probabilistic safe sets \cite{abate2008probabilistic} quantify whether or not constraint violation occurs with some probability. Indeed, let $\epsilon \in [0,1]$ be a maximum tolerable probability of constraint violation. Choose $\alpha = 1$, $r = \epsilon$, and $g_K = I_{\Bar{K}}$, where $I_{\Bar{K}}(x) = 1$ if $x \notin K$ and $I_{\Bar{K}}(x) = 0$ if $x \in K$. Then, the $(1, \epsilon)$-risk-sensitive safe set is
\begin{equation}
    \mathcal{S}_1^\epsilon = \left\{ x \in S : \inf_{\pi \in \Pi} E_x^\pi\Big( \max_{t = 0,1,\dots,T} I_{\Bar{K}}\left(X_t\right) \Big) \leq \epsilon \right\},
\end{equation}
which is the \emph{maximal probabilistic safe set} at the $\epsilon$-safety level \cite{abate2008probabilistic} for the system of Sec. \ref{theory}. (Ref. \cite{abate2008probabilistic} considers discrete-time stochastic hybrid systems that evolve under Markov policies.)

Risk-sensitive safe sets indicate higher degrees of safety as $\alpha$ decreases and $r$ decreases. We state this fact formally next.
\begin{lemma}\label{mylemma2}
Suppose that $1 \geq \alpha_1 \geq \alpha_2 > 0$ and $r_1 \geq r_2$. Then, $\mathcal{S}_{\alpha_2}^{r_2} \subseteq \mathcal{S}_{\alpha_1}^{r_1}$. If $\pi \in \Pi$, then $\mathcal{S}_{\alpha_2}^{r_2,\pi} \subseteq \mathcal{S}_{\alpha_1}^{r_1,\pi}$.
\end{lemma}
\begin{proof}
Let $x \in S$ and $\pi \in \Pi$. Since $1 \geq \alpha_1 \geq \alpha_2 > 0$ and $G \in L^1(\Omega,\mathcal{B}(\Omega),P_x^\pi)$,
$\text{CVaR}_{\alpha_2,x}^\pi(G) \geq \text{CVaR}_{\alpha_1,x}^\pi(G)$.
Since $G = \max_{t = 0,1,\dots,T}  g_K(X_t)$ and $g_K$ is bounded, there exists a $b \in \mathbb{R}$ such that $G(\omega) \geq b$ for almost every $\omega \in \Omega$. Since CVaR is monotonic and $b \in \mathbb{R}$,
    $\text{CVaR}_{\alpha_1,x}^\pi(G) \geq b$.
Take the infimum over $\pi \in \Pi$ to obtain
    $\inf_{\pi \in \Pi} \text{CVaR}_{\alpha_2,x}^\pi(G) \geq \inf_{\pi \in \Pi} \text{CVaR}_{\alpha_1,x}^\pi(G) \geq b$,
which holds for any $x \in S$. Now, suppose $x \in \mathcal{S}_{\alpha_2}^{r_2}$. Then,
$r_2 \geq \inf_{\pi \in \Pi} \text{CVaR}_{\alpha_2,x}^\pi(G) \geq \inf_{\pi \in \Pi} \text{CVaR}_{\alpha_1,x}^\pi(G)$.
Since $r_1 \geq r_2$, we have 
$r_1 \geq \inf_{\pi \in \Pi} \text{CVaR}_{\alpha_1,x}^\pi(G)$,
which shows that $x \in \mathcal{S}_{\alpha_1}^{r_1}$. 
The proof for the last statement is similar.
\end{proof}

The risk-sensitive safe set $\mathcal{S}_{\alpha}^r$ specifies that the $\text{CVaR}_{\alpha}$ of the worst constraint violation of the state trajectory must be below a given threshold. In contrast, the safe set in \cite{samuelson2018safety} specifies that for each $t$ the $\text{CVaR}_{\alpha}$ of the constraint violation of the state at time $t$ must be below a given threshold. \textcolor{black}{Hence, $\mathcal{S}_{\alpha}^r$ assesses the risk of the entire trajectory, whereas the safe set in \cite{samuelson2018safety} is concerned with the risk of each state in the trajectory separately. A specification that assesses the risk of the entire trajectory may be preferable in certain applications because this approach treats the trajectory as a unified entity representing the behavior of a control system.}
%
\subsection{Under-Approximation Method}
Risk-sensitive safe sets are well-motivated but difficult to compute due to the presence of the $\text{CVaR}$ and the maximum. Before presenting our approach to estimate risk-sensitive safe sets, we describe related methods in further detail.

Several methods in the literature apply state-space augmentation techniques to estimate the risk of a random cost incurred by a MDP.\footnote{An approach that does not require state-space augmentation is to evaluate a cumulative cost via a composition of risk functionals \cite{ruszczynski2010risk}. We take inspiration from this idea in Sec. \ref{sectheory}.} B\"{a}uerle and Ott use dynamic programming (DP) to minimize the CVaR of a sum of stage costs by defining an augmented state space \cite{bauerleott}. The range of the second state is $[0, \text{ess}\sup\sum_{t=0}^{T} C_t]$, where $C_t$ is the stage cost at time $t$ \cite[Remark 5.1]{bauerleott}. This state-space augmentation approach has been extended to optimize certainty equivalent risk functionals for MDPs \cite{bauerlerieder}. A certainty equivalent approximates the sum of the expectation and a function of the variance under particular conditions \cite{bauerlerieder}, and more generally, characterizes risk aversion in terms of functions of moments. However, CVaR provides a quantitative characterization of risk aversion by penalizing a random cost in a given fraction of the worst cases. 

Chow et al. proposed a DP algorithm to minimize approximately the CVaR of a cumulative cost via state-space augmentation, where the additional state ranges from 0 to 1 \cite{chow2015risk}. This approach is expected to be more tractable than the approach in \cite{bauerleott}; compare the ranges of the additional states. However, it is not known if the algorithm in \cite{chow2015risk} provides an upper bound or a lower bound to the solution to a CVaR-MDP problem. The algorithm in \cite{chow2015risk} is based on a CVaR Decomposition Theorem \cite[Thm. 6]{pflug2016timeEuro} \cite[Thm. 21, Lemma 22]{pflug2016timeMath}, which requires knowledge of the history of a stochastic process. How to remove the history dependence and apply the Decomposition Theorem to derive the algorithm in \cite{chow2015risk} is still an open research question.

The algorithms invented by \cite{bauerleott, chow2015risk} aim to minimize the CVaR of a \emph{cumulative} cost subject to the dynamics of a MDP. The algorithm proposed by \cite{pflug2016timeEuro} aims to minimize the CVaR of a more general cost (not necessarily a sum) but is history-dependent, which limits its computational tractability. The proof of the DP algorithm in \cite{pflug2016timeEuro} requires an exchange between an essential supremum and an expectation, whose validity in multi-stage settings for MDPs with Borel state and control spaces is not known.

Here, we propose a method to provide tractable, theoretically guaranteed under-approximations to risk-sensitive safe sets, which we define via CVaR. We focus on CVaR due to its \emph{quantitative characterization of risk aversion} and since we aim to assess the degree of safety of a control system in terms of rarer, higher-consequence outcomes. In contrast, a certainty equivalent assesses risk in terms of functions of variance and other moments. In particular, variance does not distinguish between rarer, higher-consequence outcomes in the upper tail and rarer, lower-consequence outcomes in the lower tail. Unlike the methods \cite{bauerleott, chow2015risk, bauerlerieder}, our method does not use state-space augmentation because this technique typically reduces computational tractability. For this reason, we do not augment the state space with the running maximum over each time period $Z_t \coloneqq \max_{i = 0,1,\dots,t} g_K(X_i)$. The range of $Z_t$ may be large since the bounds of $g_K$ may be large. Instead of using state-space augmentation to handle the CVaR and the maximum, we use a scaled expectation to upper bound the CVaR and a log-sum-exponential function to upper bound the maximum, $G \coloneqq \max_{t=0,1,\dots,T}g_K(X_t)$. Our first main result is below.
\begin{theorem}[Upper Bound for CVaR of $G$]\label{thm11}
For any $\pi \in \Pi$, $x \in S$, $\alpha \in (0,1]$, and $\gamma \geq 1$, it holds that
\begin{equation}\label{thm11eq}
    W_\alpha(x,\pi) \coloneqq \text{CVaR}_{\alpha,x}^\pi(G) \hspace{-0.5mm}\leq\hspace{-0.5mm} \textstyle\frac{1}{\gamma}\log\hspace{-0.5mm}\left(\hspace{-0.5mm}\textstyle\frac{1}{\alpha} E_x^\pi\Big(\hspace{-0.5mm}\sum_{t=0}^T e^{\gamma g_K(X_t)} \Big)\hspace{-0.5mm} \right)\hspace{-0.5mm}.
\end{equation}
\end{theorem}
The quantity $W_\alpha(x,\pi)$ represents the maximum extent of constraint violation of the state trajectory, averaged over the $\alpha \cdot 100$\% worst cases, when the system uses the policy $\pi$ and starts from the state $x$. The right-hand-side of \eqref{thm11eq} can be estimated more readily than $W_\alpha(x,\pi)$ for small $\alpha$ and provides a conservative approximation to $W_\alpha(x,\pi)$. If $\alpha$ is small, more samples of $G$ are required to estimate $W_\alpha(x,\pi) = \text{CVaR}_{\alpha,x}^\pi(G)$ since small $\alpha$ corresponds to rarer larger realizations of $G$. (We are more interested in using small $\alpha$ for safety-critical applications.) Theorem \ref{thm11} is powerful because it can be used to estimate the performance of \emph{any} control policy $\pi \in \Pi$ with respect to $W_\alpha(x,\pi)$. Policies may be designed for different objectives, e.g., efficiency in power or fuel consumption, robustness to bounded adversarial disturbances, robustness to bounded non-linearities, etc. It may be beneficial to estimate their performance with respect to a risk-sensitive safety criterion, such as $W_\alpha(x,\pi)$, efficiently. The proof of Theorem \ref{thm11} requires two lemmas.
\begin{lemma}[CVaR-Expectation Inequality]\label{lemmalemma3}
Let $(\Omega,\mathcal{F},\mu)$ be a probability space, $Y \in L^1(\Omega,\mathcal{F},\mu)$ such that $Y \geq 0$ a.e. w.r.t. $\mu$, and $\alpha \in (0,1]$. Then,
   $ \text{CVaR}_\alpha(Y) \leq \textstyle\frac{1}{\alpha}E(Y)$.
\end{lemma} 

A version of the inequality is stated without proof in \cite{shapiro2012}. We provide a short proof below.

\begin{proof}
Start from the CVaR definition \eqref{cvardef}, and select $s = 0$. Then,
$\text{CVaR}_{\alpha}(Y) \leq \textstyle\frac{1}{\alpha} E(\max(Y,0))$.
Since $Y \geq 0$ a.e., $\max(Y,0) = Y$ a.e., so
$\text{CVaR}_{\alpha}(Y) \leq \textstyle\frac{1}{\alpha}E(Y)$.
\end{proof}

Lemma \ref{lemmalemma3} provides an upper bound for CVaR in terms of the expectation and the risk-sensitivity level $\alpha$ when non-negative random variables are evaluated. In addition to Lemma \ref{lemmalemma3}, the proof of Theorem \ref{thm11} requires the following result, which relates the CVaR of the logarithm to the logarithm of the CVaR.
\begin{lemma}[CVaR-Log Inequality]\label{lemmalemma4}
Let $\alpha \in (0,1]$ and $Y \in L^\infty(\Omega,\mathcal{F},\mu)$. Suppose that there are real numbers $\bar{b} \geq \underline{b} > 0$ such that $\bar{b} \geq Y(\omega) \geq \underline{b}$ for every $\omega \in \Omega$. Then,
    $\text{CVaR}_{\alpha}(\log (Y)) \leq \log(\text{CVaR}_{\alpha}(Y))$.
\end{lemma} 

\begin{proof} 
%
Let $\alpha \in (0,1]$ and $\xi \in \mathcal{A}_\alpha$ \eqref{12c}. Define
    $\textstyle\mu_{\xi}(B) \coloneqq \int_B \xi \mathrm{d}\mu$, where $B \in \mathcal{F}$.
$(\Omega,\mathcal{F},\mu_{\xi})$ is a probability space, and $\int_{\Omega} Y \mathrm{d}\mu_\xi$ is finite. 
View $Y$ as a random variable on $(\Omega,\mathcal{F},\mu_{\xi})$.
It holds that $Y(\omega) \in (0,\infty)$ for all $\omega \in \Omega$, and 
$-\log$ is a convex function from $(0,\infty)$ to $\mathbb{R}$. 
Thus, by Jensen's Inequality, 
    $\textstyle\int_\Omega -\log(Y) \; \mathrm{d}\mu_{\xi}  \geq -\log\left(\textstyle\int_\Omega Y \; \mathrm{d}\mu_{\xi}\right)$.
Moreover, since $Y$ is non-negative and bounded everywhere, $\xi$ is non-negative and bounded a.e., and by using the definition of $\mu_{\xi}$, it follows that
 \begin{equation}\label{37}
     \textstyle\log\left(\int_\Omega Y  \xi \; \mathrm{d}\mu\right) \geq \int_\Omega \log(Y)  \xi \; \mathrm{d}\mu.
\end{equation}
Since $\xi \in \mathcal{A}_\alpha$ is arbitrary in the analysis above, the inequality \eqref{37} holds for all $\xi \in \mathcal{A}_\alpha$. In addition, we have
$\text{CVaR}_\alpha(Y) = \sup_{\xi \in \mathcal{A}_\alpha} \int_\Omega Y \xi \mathrm{d}\mu$ by \eqref{12}, $\text{CVaR}_\alpha(Y) \in \mathbb{R}$ because $Y \in L^1(\Omega,\mathcal{F},\mu)$, and $\int_\Omega Y  \mathrm{d}\mu_{\xi} = \int_\Omega Y  \xi \mathrm{d}\mu \geq \underline{b} > 0$ for all $\xi \in \mathcal{A}_\alpha$. 
Thus,
    $\textstyle\log\big(\text{CVaR}_\alpha(Y)\big) = \log\left(\sup_{\xi \in \mathcal{A}_\alpha} \int_\Omega Y  \xi \mathrm{d}\mu\right) \in \mathbb{R}$.
Since the natural logarithm is increasing,
\begin{equation}\label{38}
\textstyle\log\big(\text{CVaR}_\alpha(Y)\big) \geq \log\left(\int_\Omega Y  \xi \; \mathrm{d}\mu\right) \; \; \; \forall \xi \in \mathcal{A}_\alpha.
\end{equation}
By \eqref{37} and \eqref{38}, it holds that 
    $\textstyle\log\big(\text{CVaR}_\alpha(Y)\big) \geq \int_\Omega \log(Y)  \xi \mathrm{d}\mu$ for all $\xi \in \mathcal{A}_\alpha$.
Since the supremum is the least upper bound, we conclude that
    $\log\big(\text{CVaR}_\alpha(Y)\big) \geq \sup_{\xi \in \mathcal{A}_\alpha} \textstyle \int_\Omega \log(Y)  \xi \mathrm{d}\mu = \text{CVaR}_\alpha\big(\log(Y)\big)$.
\end{proof}

We use Lemma \ref{lemmalemma3} and Lemma \ref{lemmalemma4} to prove Theorem \ref{thm11}.

\hspace{-7mm}\begin{proof}[Theorem \ref{thm11}]
Note the log-sum-exp approximation for the maximum \cite[Sec. 3.1.5, p. 72]{boyd2004convex}:
If $y \in \mathbb{R}^p$ and $\gamma \geq 1$, then
\begin{equation}\label{42}
\textcolor{black}{\max_{i=1,\dots,p} y_i \overset{(a)}{\leq} {\textstyle\frac{1}{\gamma}}\log\left(\textstyle\sum_{i=1}^p e^{\gamma y_i}\right)} \leq \max_{i=1,\dots,p} y_i + {\textstyle\frac{\log(p)}{\gamma}}.
\end{equation}

Let $\pi \in \Pi$, $x \in S$, $\alpha \in (0,1]$, and $\gamma \geq 1$. Recall that $G = \underset{t=0,1,\dots,T}{\text{max}}\; g_K(X_t) \in L^\infty(\Omega,\mathcal{B}(\Omega),P_x^\pi)$, where we have presented $\Omega$ and $P_x^\pi$ at the start of Sec. \ref{theory}. 
Since $g_K$ is $\mathbb{R}$-valued, 
\begin{equation}
    Y(\omega) \coloneqq \textstyle \sum_{t=0}^T e^{\gamma g_K(X_t(\omega))} > 0 \;\;\; \forall \omega \in \Omega.
\end{equation}
Since $g_K$ is bounded and $Y$ is a sum of finitely many exponential functions of $g_K$, there exist real numbers $\bar{b} \geq \underline{b} > 0$ such that 
    $\bar{b} \geq Y(\omega) \geq \underline{b}$
for every $\omega \in \Omega$. 
It follows that $Y \in L^\infty(\Omega,\mathcal{B}(\Omega),P_x^\pi)$ satisfies the assumptions of Lemma \ref{lemmalemma4}, and thus,
\begin{equation}\label{45}
        \text{CVaR}_{\alpha,x}^\pi(\log (Y)) \leq \log(\text{CVaR}_{\alpha,x}^\pi(Y)).
\end{equation}
By the inequality $(a)$ in \eqref{42} and by the definitions of $G$ and $Y$, the inequality 
    $G \leq {\textstyle\frac{1}{\gamma}}\log\left(\textstyle\sum_{t=0}^T e^{\gamma g_K(X_t)}\right) = {\textstyle\frac{1}{\gamma}}\log(Y)$
holds a.e. w.r.t. $P_x^\pi$. Since CVaR is monotonic and positively homogeneous, and since $\textstyle\frac{1}{\gamma} > 0$,
\begin{equation}\label{47}
    \text{CVaR}_{\alpha,x}^\pi ( G ) \leq \text{CVaR}_{\alpha,x}^\pi\big({\textstyle\frac{1}{\gamma}}\log(Y) \big) = {\textstyle\frac{1}{\gamma}}\text{CVaR}_{\alpha,x}^\pi(\log(Y) ).
\end{equation}
We use \eqref{45} and \eqref{47} to find that
\begin{equation}\label{48}
    \text{CVaR}_{\alpha,x}^\pi ( G ) \leq {\textstyle\frac{1}{\gamma}}\log(\text{CVaR}_{\alpha,x}^\pi(Y)).
\end{equation}
Note that $\text{CVaR}_{\alpha,x}^\pi(Y) \in \mathbb{R}$ such that $\text{CVaR}_{\alpha,x}^\pi(Y) > 0$. Indeed,
$Y \in L^\infty(\Omega,\mathcal{B}(\Omega),P_x^\pi)$ and so is also an element of $L^1(\Omega,\mathcal{B}(\Omega),P_x^\pi)$, hence $\text{CVaR}_{\alpha,x}^\pi(Y) \in \mathbb{R}$. $Y$ is bounded everywhere, and in particular, from below by a real number $\underline{b} >0$. Therefore, $\text{CVaR}_{\alpha,x}^\pi(Y) \geq \text{CVaR}_{\alpha,x}^\pi(\underline{b}) = \underline{b} > 0$. Consequently, $\log(\text{CVaR}_{\alpha,x}^\pi(Y)) \in \mathbb{R}$. In addition, the assumptions of Lemma \ref{lemmalemma3} are satisfied, and therefore,
\begin{equation}\label{499}
    \text{CVaR}_{\alpha,x}^\pi(Y) \leq \textstyle\frac{1}{\alpha}E_x^\pi(Y).
\end{equation}
Use \eqref{48}, \eqref{499}, and $\log$ being increasing to derive that
    $\text{CVaR}_{\alpha,x}^\pi ( G ) \leq {\textstyle\frac{1}{\gamma}}\log(\text{CVaR}_{\alpha,x}^\pi(Y)) \leq {\textstyle\frac{1}{\gamma}}\log\left(\textstyle\frac{1}{\alpha}E_x^\pi(Y)\right)$.
\end{proof}

We use the conclusion of Theorem \ref{thm11} to define particular subsets of the state space. First, we call these sets approximations, and then, we prove that they are under-approximations to risk-sensitive safe sets in Theorem \ref{thm22}.
\begin{definition}[Approximations to Risk-Sensitive Safe Sets]\label{approxsets}
Let $\alpha \in (0,1]$, $r \in \mathbb{R}$, and $\gamma \geq 1$ be given. The \emph{$(\alpha,r,\gamma)$-approximation set for a given policy $\pi \in \Pi$} is defined by
\begin{equation}\label{myUgivenpolicy}\begin{aligned}
\mathcal{U}_{\alpha,\gamma}^{r,\pi} & \coloneqq \left\{x \in S : \textstyle\frac{1}{\alpha}E_{x}^\pi\Big(\sum_{t=0}^T e^{\gamma g_K(X_t)}\Big) \leq e^{\gamma r} \right\}.
\end{aligned}\end{equation}
The \emph{$(\alpha,r,\gamma)$-approximation set} is defined by
\begin{equation}\label{myU}\begin{aligned}
\mathcal{U}_{\alpha,\gamma}^{r} & \coloneqq \left\{x \in S : \inf_{\pi \in \Pi} \textstyle\frac{1}{\alpha}E_{x}^\pi\Big(\sum_{t=0}^T e^{\gamma g_K(X_t)}\Big) \leq e^{\gamma r}\right\}.
\end{aligned}\end{equation}
\end{definition}
We denote the infimum in \eqref{myU} by
\begin{equation}\label{Jalphastar}
J_{\alpha,\gamma}^*(x) \coloneqq \inf_{\pi \in \Pi} J_{\alpha,\gamma}(x,\pi) \coloneqq \inf_{\pi \in \Pi} \textstyle\frac{1}{\alpha}E_{x}^\pi\Big(\sum_{t=0}^T e^{\gamma g_K(X_t)}\Big),
\end{equation}
where $\Pi$ is the set of randomized history-dependent policies, which also includes deterministic Markov policies. Estimating $J_{\alpha,\gamma}^*$ is the critical step for estimating the sets $\mathcal{U}_{\alpha,\gamma}^{r}$. The problem of estimating $J_{\alpha,\gamma}^*$ is a Markov decision process problem. Thus, $J_{\alpha,\gamma}^*$ and a deterministic Markov policy $\pi_\gamma \in \Pi$ such that $J_{\alpha,\gamma}(x,\pi_\gamma) = J_{\alpha,\gamma}^*(x)$ for all $x \in S$ can be computed via dynamic programming, in principle, if a measurable selection condition holds.\footnote{Measurable selection conditions, e.g., see \cite[Chapter 3.3]{hernandez2012discrete} or \cite{bertshreve2004}, are commonly invoked to guarantee the existence of a policy that optimizes or nearly optimizes an expected cumulative cost subject to a MDP.\label{footnote11}} Therefore, for a fixed $\gamma \geq 1$, an algorithm to estimate $\{J_{\alpha,\gamma}^* : \alpha \in \Lambda\}$, where $\Lambda \subseteq (0,1]$ is a family of risk-sensitivity levels, exists and is tractable. The next theorem shows that the sets in Definition \ref{approxsets} are under-approximations to risk-sensitive safe sets (Definition \ref{risksafedef}).
\begin{theorem} \emph{(Under-Approximations to Risk-Sensitive Safe Sets)} \label{thm22}
Let $\alpha \in (0,1]$, $r \in \mathbb{R}$, and $\gamma \geq 1$. For any policy $\pi \in \Pi$, it holds that
\begin{equation}\label{subsetgivenpolicy}
    \mathcal{U}_{\alpha,\gamma}^{r,\pi} \subseteq \mathcal{S}_{\alpha}^{r,\pi},
\end{equation}
where $\mathcal{U}_{\alpha,\gamma}^{r,\pi}$ is defined by \eqref{myUgivenpolicy} and $\mathcal{S}_{\alpha}^{r,\pi}$ is defined by \eqref{mySgivenpolicy}. Moreover, the $(\alpha,r,\gamma)$-approximation set is a subset of the $(\alpha,r)$-risk-sensitive safe set, i.e.,
\begin{equation}\label{subsetUS}
    \mathcal{U}_{\alpha,\gamma}^{r} \subseteq \mathcal{S}_{\alpha}^{r},
\end{equation}
where $\mathcal{U}_{\alpha,\gamma}^{r}$ is defined by \eqref{myU} and $\mathcal{S}_{\alpha}^{r}$ is defined by \eqref{myS}.
\end{theorem}
\begin{proof}
Eq. \eqref{subsetgivenpolicy} follows from Theorem \ref{thm11}.  Let $\alpha \in (0,1]$, $r \in \mathbb{R}$, $\gamma \geq 1$, and $\pi \in \Pi$ be given. Let $x \in \mathcal{U}_{\alpha,\gamma}^{r,\pi}$. Then,
\begin{equation}
    \textstyle\frac{1}{\alpha}E_{x}^\pi\Big(\sum_{t=0}^T e^{\gamma g_K(X_t)}\Big) \leq e^{\gamma r},
\end{equation}
where the left-hand-side is bounded below by a positive real number since $Y \coloneqq \sum_{t=0}^T e^{\gamma g_K(X_t)}$ is as well. 
It follows that $\log\left(\textstyle\frac{1}{\alpha}E_{x}^\pi\left(\sum_{t=0}^T e^{\gamma g_K(X_t)}\right)\right)$ is finite. Since the natural logarithm is increasing and $\gamma\geq 1$, we have
\begin{equation}\label{57}
    \textstyle\frac{1}{\gamma}\log\left(\textstyle\frac{1}{\alpha}E_{x}^\pi\Big(\sum_{t=0}^T e^{\gamma g_K(X_t)}\Big)\right) \leq r.
\end{equation}
By Theorem \ref{thm11}, it holds that
\begin{equation}\label{58}
    \text{CVaR}_{\alpha,x}^\pi(G) \leq \textstyle\frac{1}{\gamma}\log\left(\textstyle\frac{1}{\alpha} E_x^\pi\Big(\sum_{t=0}^T e^{\gamma g_K(X_t)} \Big) \right).
\end{equation}
Combine \eqref{57} and \eqref{58} to find that
    $\text{CVaR}_{\alpha,x}^\pi(G) \leq r$,
which shows that $x \in \mathcal{S}_{\alpha}^{r,\pi}$ and proves \eqref{subsetgivenpolicy}. Now, to prove \eqref{subsetUS}, let $x \in \mathcal{U}_{\alpha,\gamma}^{r}$, which implies that
\begin{equation}\label{600}
    \inf_{\pi \in \Pi} \textstyle\frac{1}{\alpha}E_{x}^\pi\Big(\sum_{t=0}^T e^{\gamma g_K(X_t)}\Big) \leq e^{\gamma r}.
\end{equation}
Let $\epsilon > 0$ be given. Since the left-hand-side of \eqref{600} is finite, there is a $\pi^\epsilon \in \Pi$ such that 
\begin{equation}\label{61}\begin{aligned}
    {\textstyle\frac{1}{\alpha}}E_x^{\pi^\epsilon}\left({\textstyle\sum_{t=0}^T}  e^{\gamma g_K(X_t)}\right) & \leq \epsilon + \inf_{\pi \in \Pi} {\textstyle\frac{1}{\alpha}}E_x^\pi\left(\textstyle\sum_{t=0}^T  e^{\gamma g_K(X_t)}\right) \\
    & \leq \epsilon + e^{\gamma r},
\end{aligned}\end{equation}
where the second line holds by \eqref{600}. 
 Note that the quantity $\log\left({\textstyle\frac{1}{\alpha}}E_x^{\pi^\epsilon}\left({\textstyle\sum_{t=0}^T}  e^{\gamma g_K(X_t)}\right)\right)$ is finite.
Take the logarithm of \eqref{61} and then divide by $\gamma \geq 1$ to obtain
\begin{equation}
    {\textstyle\frac{1}{\gamma}}\log\left({\textstyle\frac{1}{\alpha}}E_x^{\pi^\epsilon}\left({\textstyle\sum_{t=0}^T}  e^{\gamma g_K(X_t)}\right)\right) \leq {\textstyle\frac{1}{\gamma}}\log\left(\epsilon + e^{\gamma r}\right).
\end{equation}
By Theorem \ref{thm11}, it holds that
\begin{equation}
    \text{CVaR}_{\alpha,x}^{\pi^\epsilon}(G) \leq \textstyle\frac{1}{\gamma}\log\left(\textstyle\frac{1}{\alpha} E_x^{\pi^\epsilon}\Big(\sum_{t=0}^T e^{\gamma g_K(X_t)} \Big) \right).
\end{equation}
Therefore,
   $ \text{CVaR}_{\alpha,x}^{\pi^\epsilon}(G) \leq {\textstyle\frac{1}{\gamma}}\log\left(\epsilon + e^{\gamma r}\right)$.
Since $\pi^\epsilon \in \Pi$, it follows that 
\begin{equation}\begin{aligned}
    W_\alpha^*(x)  \coloneqq \inf_{\pi \in \Pi} \text{CVaR}_{\alpha,x}^{\pi}(G)  \leq  \text{CVaR}_{\alpha,x}^{\pi^\epsilon}(G).
\end{aligned}\end{equation}
Consequently, we have 
\begin{equation}
    W_\alpha^*(x) \leq \textstyle\frac{1}{\gamma}\log\left(\epsilon + e^{\gamma r}\right).
\end{equation}
This analysis holds for any $\epsilon > 0$. Let $\epsilon \rightarrow 0$, and use the continuity of the logarithm to obtain
\begin{equation}\begin{aligned}
    W_{\alpha}^*(x)  \leq \lim_{\epsilon \rightarrow 0} {\textstyle\frac{1}{\gamma}}\log\left(\epsilon + e^{\gamma r} \right)
    = {\textstyle\frac{1}{\gamma}}\log\left(\lim_{\epsilon \rightarrow 0}\epsilon + e^{\gamma r} \right) =  r.
\end{aligned}\end{equation}
Since $W_{\alpha}^*(x) \leq r$, we conclude that $x \in  \mathcal{S}_\alpha^r$. Since any $x \in \mathcal{U}_{\alpha,\gamma}^r$ is also an element of $\mathcal{S}_\alpha^r$, it holds that $\mathcal{U}_{\alpha,\gamma}^r \subseteq \mathcal{S}_\alpha^r$.
\end{proof}

Since we have shown that $\mathcal{U}_{\alpha,\gamma}^{r,\pi}$ and $\mathcal{U}_{\alpha,\gamma}^{r}$ are subsets of the risk-sensitive safe sets, $\mathcal{S}_{\alpha}^{r,\pi}$ and $\mathcal{S}_{\alpha}^{r}$, respectively, we now refer to $\mathcal{U}_{\alpha,\gamma}^{r,\pi}$ and $\mathcal{U}_{\alpha,\gamma}^{r}$ as \emph{under-approximations}.

\begin{remark}[Assessment of Approximation Errors]
Three approximations are required for the proof above. First, we use a soft-maximum, under which we have
\begin{equation}\label{44}
    0 \leq \textstyle \frac{1}{\gamma} \text{CVaR}_{\alpha,x}^\pi(\log(Y)) -\text{CVaR}_{\alpha,x}^\pi(G) \leq \frac{\log(T+1)}{\gamma},
\end{equation}
where $Y = \sum_{t=0}^T e^{\gamma g_K(X_t)}$, and there are positive constants $\underline{b}$ and $\bar{b}$ (which depend on $T$, $\gamma$, and the bounds of $g_K$) such that $Y \in [\underline{b}, \bar{b}]$ everywhere. The inequality \eqref{44} implies an improved approximation with larger values of $\gamma$ or smaller values of $T$. However, since it is not feasible to optimize $\frac{1}{\gamma} \text{CVaR}_{\alpha,x}^\pi(\log(Y))$ directly, our next step is to leverage the CVaR-log inequality provided by Lemma \ref{lemmalemma4}. The associated error is given by
\begin{equation}
  \eta_{\alpha,x}^{\pi,\gamma} \coloneqq \textstyle \frac{1}{\gamma}\log(\text{CVaR}_{\alpha,x}^{\pi}(Y)) - \frac{1}{\gamma}\text{CVaR}_{\alpha,x}^{\pi}(\log(Y)) \geq 0. 
\end{equation}
Since the range of $Y$ is $[\underline{b}, \bar{b}]$, it follows that $\eta_{\alpha,x}^{\pi,\gamma} \leq \frac{1}{\gamma}\log(\bar{b}/\underline{b})$. Therefore, we anticipate a smaller error $\eta_{\alpha,x}^{\pi,\gamma}$ when $Y$ has a smaller range, which occurs when $T$ is smaller, for example.

The last approximation is $\log(\textnormal{CVaR}_{\alpha,x}^\pi(Y)) \leq \log\big(\frac{1}{\alpha}E_x^\pi(Y)\big)$, which of course is poor as $\alpha \rightarrow 0$. However, for a fixed $\alpha \in (0,1)$, we anticipate that this approximation performs well when $P_x^\pi$ has a fat (upper) tail, which we state formally in the following lemma.
\end{remark}
\begin{lemma}[Tightness of $\log(\textnormal{CVaR}_{\alpha}(Y)) \leq \log(\frac{1}{\alpha}E(Y))$]\label{tightnesslemma}
Assume the conditions of Lemma \ref{lemmalemma4}, and let $\alpha \in (0,1)$. Suppose that for some finite $m > 0$, it holds that
\begin{equation}\label{49}
   0 <  m  \int_{0}^{1-\alpha} \text{VaR}_{1-p}(Y) \; \mathrm{d}p \leq \int_{1-\alpha}^{1} \text{VaR}_{1-p}(Y) \; \mathrm{d}p.
\end{equation}
Then,
  $\textstyle  0 \leq \log\big(\frac{1}{\alpha}E(Y)\big) - \log(\text{CVaR}_{\alpha}(Y)) \leq \log\big(\frac{1}{m} + 1\big)$.
\end{lemma}
\begin{remark}[Fat tail condition \eqref{49}]
The second inequality in \eqref{49} means that the cumulative VaR in the upper $\alpha$-fraction of the distribution of $Y$, $\int_{1-\alpha}^{1} \text{VaR}_{1-p}(Y) \mathrm{d}p$, is at least $m$ times greater than the cumulative VaR in the lower $(1-\alpha)$-fraction of the distribution of $Y$, $\int_{0}^{1-\alpha} \text{VaR}_{1-p}(Y) \mathrm{d}p$. The maximum value of $m$ that satisfies \eqref{49} is $\hat{m} = \frac{\int_{1-\alpha}^{1} \text{VaR}_{1-p}(Y) \mathrm{d}p}{ \int_{0}^{1-\alpha} \text{VaR}_{1-p}(Y) \mathrm{d}p}$, which gives a measure of tail ``fatness.'' For example, if the distribution of $Y$ is a standard log-normal with parameters $\mu = 0$ and $\sigma = 1$, and if $\alpha = 0.05$, then numerical integration yields $\hat{m} \approx \frac{0.42}{1.2} \approx 0.35.$ If $\sigma$ is increased to 2 under the same conditions, then $\hat{m} \approx \frac{4.7}{2.7} \approx 1.7$.
\end{remark} 

Next, we prove Lemma \ref{tightnesslemma}.\\
\begin{proof}[Lemma \ref{tightnesslemma}]
The representation of CVaR in \eqref{2b} and the inequality \eqref{49} imply that
\begin{equation}
  \frac{1}{\alpha}\int_{0}^{1-\alpha} \text{VaR}_{1-p}(Y) \mathrm{d}p  \leq \frac{\text{CVaR}_{\alpha}(Y)}{m}.
\end{equation}
The expectation and the VaR are related by $E(Y) = \text{CVaR}_{1}(Y) = \int_{0}^{1} \text{VaR}_{1-p}(Y) \mathrm{d}p$. It follows that
   $\textstyle \frac{1}{\alpha}E(Y) \leq \left(\frac{1}{m} + 1 \right)\text{CVaR}_{\alpha}(Y)$.
From this and Lemma \ref{lemmalemma3}, we have
\begin{equation}\label{5252}
  \textstyle \text{CVaR}_{\alpha}(Y) \leq \frac{1}{\alpha}E(Y) \leq  \left(\frac{1}{m} + 1 \right)\text{CVaR}_{\alpha}(Y).
\end{equation}
Then, take the logarithm of \eqref{5252} and subtract $\log(\text{CVaR}_{\alpha}(Y)) \in \mathbb{R}$ to complete the derivation.
\end{proof}

From Theorem \ref{thm22}, we obtain tractable under-approximations to risk-sensitive safe sets. In practice, one selects $\gamma \geq 1$ manually and then estimates $J_{\alpha,\gamma}^*$ \eqref{Jalphastar} for a family of risk-sensitivity levels. For a fixed $\gamma$, only \emph{one} MDP problem on the original state space needs to be solved for any family of risk-sensitivity levels because $J_{\alpha,\gamma}^*$ is a standard MDP problem scaled by $\alpha$. In Sec. \ref{num}, which presents numerical examples, we take one approach to choose a suitable value of $\gamma$ manually by visual inspection. Before proceeding to the numerical examples, we present one additional theoretical contribution.
\section{Toward a Parameter-Independent Safety Analysis Framework}\label{sectheory}
Previously, we have defined risk-sensitive safe sets in terms of the CVaR of a maximum random cost. However, this risk-sensitive safety criterion is difficult to optimize exactly without using state-space augmentation, which motivated us to derive a parameter-dependent upper bound. One may wonder whether there is another coherent risk functional (ideally related to CVaR) that admits an upper bound, which can be computed via DP on the original state space without an additional parameter that requires tuning. The answer is indeed positive, as presented below.

\begin{definition}[Proposed Risk Functional]\label{newrisk}
Let $\alpha \in (0,1]$, $x \in S$, $\pi \in \Pi$, and $Y \in L^\infty(\Omega,\mathcal{B}(\Omega),P_x^\pi)$ be given. Let $\mathcal{D}_{\alpha}$ be a set of tuples of densities. Each tuple $\zeta \in \mathcal{D}_{\alpha}$ takes the form $\zeta = (\xi_0,\xi_1,\dots,\xi_{T-1})$, where the properties of the densities follow. For each $t$, $\xi_t(\cdot|\cdot,\cdot) : S \times S \times A \rightarrow \mathbb{R}$ is Borel measurable, and for every $(x,u) \in S \times A$, it holds that $\xi_t(\cdot|x,u) \in \mathcal{R}_\alpha(x,u)$. Here, $\mathcal{R}_\alpha(x,u)$ is the set of Borel-measurable functions of the form $\nu : S \rightarrow \mathbb{R}$ such that $\nu \in [0, \alpha^{-1/T}]$ a.e. w.r.t. $Q(\cdot|x,u)$ and $\int_S \nu \hspace{.5mm} \mathrm{d}Q(\cdot|x,u) = 1$. We define $\rho_{\alpha,x}^\pi(Y)$ by
\begin{equation}\label{544}
\rho_{\alpha,x}^\pi(Y) \coloneqq \underset{ (\xi_0,\xi_1,\dots,\xi_{T-1})\in \mathcal{D}_{\alpha}}{\sup}  \int_{\Omega} \hspace{-.5mm} Y \hspace{-.5mm} \textcolor{black}{\prod\limits_{t=0}^{T-1} \xi_t(x_{t+1}|x_t,u_t)} \hspace{.5mm} \mathrm{d}P_x^\pi.
\end{equation}
\end{definition}
\begin{remark}[Interpretation for $\mathcal{R}_\alpha(x,u)$]
$\mathcal{R}_\alpha(x,u)$ is related to the set of densities in the CVaR representation given by \eqref{12}. If the probability space is $(S, \mathcal{B}(S), Q(\cdot|x,u))$, then $\mathcal{A}_{\alpha'} = \mathcal{R}_\alpha(x,u)$, where $\alpha' = \alpha^{1/T}$.
\end{remark}

\begin{remark}[Interpretation for $\rho_{\alpha,x}^\pi$]
Although we do not yet have an exact interpretation for $\rho_{\alpha,x}^\pi$, we provide a preliminary interpretation here. The quantity $\rho_{\alpha,x}^\pi(Y)$ is a distributionally robust expectation of $Y$, such that an uncertainty $\xi_t$ perturbs the system's nominal transition law $Q$ at each time $t$. $\xi_t$ may depend on the current time, state, and control. Moreover, $\rho_{\alpha,x}^\pi(Y)$ strikes a balance between the expectation and CVaR, as formalized below.
\end{remark}
\begin{lemma}[Coherence of $\rho_{\alpha,x}^\pi$, relation to CVaR]\label{coherent}
The risk functional $\rho_{\alpha,x}^\pi : L^\infty(\Omega,\mathcal{B}(\Omega),P_x^\pi) \rightarrow \mathbb{R}$ is coherent. In addition, for any $Y \in L^\infty(\Omega,\mathcal{B}(\Omega),P_x^\pi)$, the inequality 
    $E_x^\pi(Y) \leq \rho_{\alpha,x}^\pi(Y) \leq \text{CVaR}_{\alpha,x}^\pi(Y)$ holds.
\end{lemma}
\begin{proof}
The first step is to verify the properties of monotonicity, subadditivity, translation equivariance, and positive homogeneity, which we omit in the interest of space. To show that $E_x^\pi(Y) \leq \rho_{\alpha,x}^\pi(Y)$, note that $\zeta = (\xi_0,\xi_1,\dots,\xi_{T-1})$ such that $\xi_t$ equals 1 for each $t$ is an element of $\mathcal{D}_\alpha$. The inequality $\rho_{\alpha,x}^\pi(Y) \leq \text{CVaR}_{\alpha,x}^\pi(Y)$ follows from \eqref{9b}--\eqref{9c}.
\end{proof}
%
%

We use the risk functional \eqref{544} to define a safe set.
\begin{definition}[$\bar{\mathcal{S}}_\alpha^{r}$-Risk-Sensitive Safe Set]\label{rss}
For any $\alpha \in (0,1]$ and $r \in \mathbb{R}$, define $\bar{\mathcal{S}}_\alpha^{r} \coloneqq \{ x \in S : \inf_{\pi \in \Pi} \rho_{\alpha,x}^\pi(Y) \leq r \}$.
\end{definition}

Definition \ref{rss} is inspired by Definition \ref{risksafedef}, and the form of $\rho_{\alpha,x}^\pi$ \eqref{544} is inspired by the representation for CVaR in \eqref{9b}--\eqref{9c}. We emphasize a key distinction. In \eqref{544}, there is a function $\xi_t$ for each $t$ that depends on the current state and control. In \eqref{9b}--\eqref{9c}, however, each function in $\mathcal{A}_{\alpha,x}^\pi$ depends on the entire history. The ``separable'' structure of \eqref{544} allows us to derive a DP algorithm on the original state space to upper bound $\inf_{\pi \in \Pi} \rho_{\alpha,x}^\pi(Y)$ without using a parameter that requires tuning. 
In this section, we make two assumptions.
\begin{assumption}[Properties of $Y$]\label{assumption0}
We consider the case when $Y \coloneqq c_T(X_T) +\sum_{t=0}^{T-1} c_t(X_t,U_t)$ is cumulative. The functions $c_t : S \times A \rightarrow \mathbb{R}$ for all $t \in \{0,1,\dots,T-1\}$ and $c_T : S \rightarrow \mathbb{R}$ are bounded and upper semi-continuous (usc).
\end{assumption}
\begin{assumption}[Continuity property of $Q$] \label{assumption3}
\textcolor{black}{The transition kernel $Q$ \eqref{mytransitionlaw} is continuous in total variation; that is, if $(x_n,u_n) \rightarrow (x,u)$, then $|Q(\cdot|x_n,u_n) -Q(\cdot|x,u) |(S) \rightarrow 0$}.
\end{assumption}
%
%
\begin{remark}[Example that satisfies Assumption \ref{assumption3}]
\textcolor{black}{Suppose that $P_D$ has a continuous non-negative density and $f$ in \eqref{mytransitionlaw} has the form $f(x,u,d) = f_1(x,u) + d \cdot f_2(x,u)$, where $W = S$ is a vector space with field $\mathbb{R}$, $f_1 : S \times A \rightarrow S$ and $f_2 : S \times A \rightarrow \mathbb{R}$ are continuous, and $f_2$ is non-zero. Then, by Scheff{\'e}'s Lemma, Assumption \ref{assumption3} is satisfied.} We note that continuity of $f$ is a typical condition in stochastic control, e.g., see \cite[p. 209]{bertshreve2004}, and requiring additional structure on the dynamics to achieve tractable algorithms is standard. For example, under some assumptions the dynamics may be decomposed into overlapping systems, to obtain conservative under-approximations to reachable sets for continuous-time, non-stochastic systems \cite{chenconf, chen2018decomp}. A mixed monotone structure has been assumed to approximate reachable sets for discrete-time non-stochastic systems, with applications to traffic safety \cite{Cooganconf, Cooganmagazine}. More broadly, additive continuous noise is a realistic assumption in many domains, e.g., additive Gaussian noise in information theory and control (classical references include \cite{whittle1981, sinopoli2004}) and additive Brownian motion in continuous-time epidemiological modeling \cite{Lef2018, Lef2020}.
\end{remark}

Boundedness and upper semi-continuity of $c_t$ for all $t$ ensures that $Y \in L^\infty(\Omega,\mathcal{B}(\Omega),P_x^\pi)$ for any $x \in S$ and $\pi \in \Pi$. Also, boundedness of $c_t$ ensures that the iterates of a DP recursion are bounded, which we use to show that a supremum over $\mathcal{R}_\alpha(x,u)$ of the form $\phi(x,u) \coloneqq  \sup_{\xi \in \mathcal{R}_{\alpha}(x,u)}  \int_{S} J \xi  \hspace{.5mm} \mathrm{d}Q(\cdot|x,u)$ is attained (Lemma \ref{attainmentlemma}, Appendix). This attainment and Assumption \ref{assumption3} together guarantee that the supremum is usc in $(x,u)$ (Lemma \ref{keypropofphi}, Appendix). 
The upper semi-continuity of the supremum permits the derivation of an upper bound for $\inf_{\pi \in \Pi} \rho_{\alpha,x}^\pi(Y)$ via DP.
 
\begin{theorem}[DP to Upper Bound $\inf_{\pi \in \Pi} \rho_{\alpha,x}^\pi(Y)$]\label{thm33}
Let Assumptions \ref{assumption0}--\ref{assumption3} hold, and let $\alpha \in (0,1]$ be given. Define 
\begin{subequations}
\begin{equation}
    J_{T}^\alpha \coloneqq c_T,
\end{equation}
and for $t = T-1,\dots,1,0$, define 
\begin{equation}
    J_{t}^\alpha(x) \coloneqq \inf_{u \in A} v_{t}^\alpha(x,u) \;\;\; \forall x \in S,
\end{equation}
where $v_{t}^\alpha \coloneqq c_t + \varphi_t^\alpha$ and
\begin{align}
    \varphi_t^\alpha(x,u)   \coloneqq  \sup_{\xi \in \mathcal{R}_\alpha(x,u)} \int_{S}   J_{t+1}^\alpha \xi \; \mathrm{d}Q( \cdot|x,u)
\end{align}
\end{subequations}
for all $(x,u) \in S \times A$. Then, $J_{t}^\alpha$ is usc and bounded for all $t = 0,1,\dots,T$. For all $\epsilon > 0$, there is a deterministic Markov policy $\pi_\epsilon^* \in \Pi$ such that $\rho_{\alpha,x}^{\pi_\epsilon^*}(Y) \leq J_{0}^\alpha(x) + \epsilon$ for all $x \in S$. In particular, ${\inf}_{\pi \in \Pi} \; \rho_{\alpha,x}^\pi(Y) \leq J_{0}^\alpha(x)$ for all $x \in S$.
%
%
\end{theorem}

A proof for Theorem \ref{thm33} is in the Appendix, where we include supporting results as well. 

Theorem 3 is exciting for two main reasons: 1) it provides a more numerically tractable way to estimate safe sets (the upper bound does not have a parameter that requires tuning, and the algorithm does not require an augmented state space); and 2) more broadly, the result initiates new avenues for tractable solutions to risk-sensitive safety analysis problems. 
\section{Numerical Examples}\label{num}
Here, we present examples of risk-sensitive safe sets and their under-approximations as in Definition \ref{risksafedef} for a temperature system and a stormwater system.\footnote{We used the Tufts Linux Research Cluster (Medford, MA) with MATLAB (The Mathworks, Inc.). Our code is available from  \url{https://github.com/risk-sensitive-reachability/IEEE-TAC-2021}.} For each example, we have chosen a value of $\gamma$ by exploring increasing integer values and then stopping the exploration when improvements in the estimates of $\mathcal{U}_{\alpha,\gamma}^r$ were no longer apparent.
\subsection{Temperature System}\label{tempsysex}
Consider a thermostatically controlled load evolving on a finite-time horizon $t = 0,1, \dots, T-1$ via a deterministic Markov policy $\pi = (\pi_0,\pi_1,\dots,\pi_{T-1})$, $$X_{t+1} = a X_t + (1-a) (b - \eta  \bar{r}  \bar{p}  \pi_t(X_t) ) + D_t.$$ This model is from \cite{yang2018dynamic, mort}. $X_t$ is the $\mathbb{R}$-valued random temperature ($^{\circ}\text{C}$) of a thermal mass at time $t$. $\pi_t(X_t)$ is the $[0,1]$-valued control at time $t$. The amount of power supplied to the system decreases as the value of the control increases from 0 to 1. $(D_0,D_1,\dots,D_{T-1})$ is a $\mathbb{R}$-valued, iid stochastic process that arises due to environmental uncertainties. We consider three discrete distributions for the disturbance process, where each distribution has a distinct skew (left skew, no skew, or right skew). In each distribution, the minimum disturbance value is $-0.5$ $^{\circ}\text{C}$, and the maximum disturbance value is 0.5 $^{\circ}\text{C}$. Table \ref{paramtemp} provides the model parameters.
%
%
\begin{table}[h]
\caption{Temperature System Parameters}
\label{table}
\setlength{\tabcolsep}{3pt}
\begin{tabular}{|p{25pt}|p{130pt}|p{55pt}|}
\hline
Symbol &  Description & Value  \\
\hline
$a$ & time delay & $e^{\frac{-\triangle \tau}{\bar{c}\bar{r}}}$ (no units) \\
$b$ & temperature shift & 32 $^{\circ}\text{C}$ \\
$\bar{c}$ & thermal capacitance & 2 $\frac{\text{kWh}}{^{\circ}\text{C}}$ \\
$\eta$ & control efficiency & 0.7 (no units) \\
$K$ & constraint set & $[20, 21]$ $^{\circ}\text{C}$ \\
$\bar{p}$ & range of energy transfer to/from thermal mass & 14 kW \\
$\bar{r}$ & thermal resistance & 2 $\frac{^{\circ}\text{C}}{\text{kW}}$ \\
$\triangle \tau$ & duration of $[t, t+1)$ & $\frac{5}{60}$ h \\
$T$ & length of discrete time horizon & 12 (= 1 h) \\
$A$ & control space & $[0,1]$ (no units)\\
$S$ & state space & $[18, 23]$ $^{\circ}\text{C}$ \\
\hline
\multicolumn{3}{p{240pt}}{h $=$ hours, kW $=$ kilowatts, $^{\circ}\text{C}$ $=$ degrees Celsius.}
\end{tabular}
\label{paramtemp}
\end{table}

We have chosen $g_K(X_t) = \max( X_t - 21, 20 - X_t)$ to quantify the extent of constraint violation of the state $X_t$ with respect to the constraint set $K = [20, 21]$ $^{\circ}\text{C}$. $K$ is a temperature range, where the state trajectory should remain inside whenever possible. For different values of $\gamma$ (see next paragraph), we have implemented classical DP with linear interpolation to estimate
\begin{equation}\label{72}
    J_\gamma^*(x) \coloneqq \inf_{\pi \in \Pi} J_\gamma(x,\pi) \coloneqq \inf_{\pi \in \Pi} E_x^\pi \hspace{-.5mm} \left(\textstyle\sum_{t=0}^T e^{\gamma g_K(X_t)}\right)
\end{equation}
and a deterministic Markov policy $\pi_\gamma \in \Pi$ such that $J_\gamma^*(x) = J_\gamma(x,\pi_\gamma)$ for all $x \in S$. 
DP on continuous state and control spaces is implemented typically via discretization and interpolation. In particular, we have discretized the set of controls $A = [0,1]$ and the set of states $S = [18, 23]$ $^{\circ}\text{C}$ uniformly at a resolution of 0.1. To improve efficiency of DP, approximate DP methods are being developed, e.g., see \cite{lygeros2018siam, bertrl}, and the references therein. While these methods are exciting, we leave investigations of their applicability to risk-sensitive safety analysis for future work. 

We have used $\gamma \in \Gamma \coloneqq \{3,4,\dots,20\}$ because for all $y \in S$ and $\gamma \in \Gamma$, the stage cost $e^{\gamma g_K(y)}$ is at most $e^{20 \cdot 2}$, a large number that a personal computer can handle. We have considered risk-sensitivity levels from nearly risk-neutral ($\alpha = 0.99$) to more risk-averse ($\alpha$ near 0). Specifically, we have chosen $\alpha \in \Lambda \coloneqq \{ 0.99, 0.05, 0.01, 0.005, 0.001 \}$. A typical risk-sensitivity level is $\alpha = 0.05$ or $\alpha = 0.01$, and we have considered smaller values of $\alpha$ as well. For $\gamma \in \Gamma$ and $\alpha \in \Lambda$, we have estimated $J^*_{\alpha,\gamma}$ \eqref{Jalphastar} by dividing our estimate of $J_\gamma^*$ \eqref{72} by $\alpha$. Let $\hat{S}$ denote the state space grid. By using our estimate of $\pi_\gamma$, we have simulated 100,000 trajectories from each initial state $x \in \hat{S}$ to generate an empirical distribution of $G \coloneqq \max_{t=0,1,\dots,T} g_K(X_t)$. Then, for each $\alpha \in \Lambda$, we have used a consistent CVaR estimator \cite[p. 300]{shapiro2009lectures} to estimate $\text{CVaR}_{\alpha,x}^{\pi_\gamma}(G)$.

\begin{figure*}[t]
\centerline{\includegraphics[width=\textwidth]{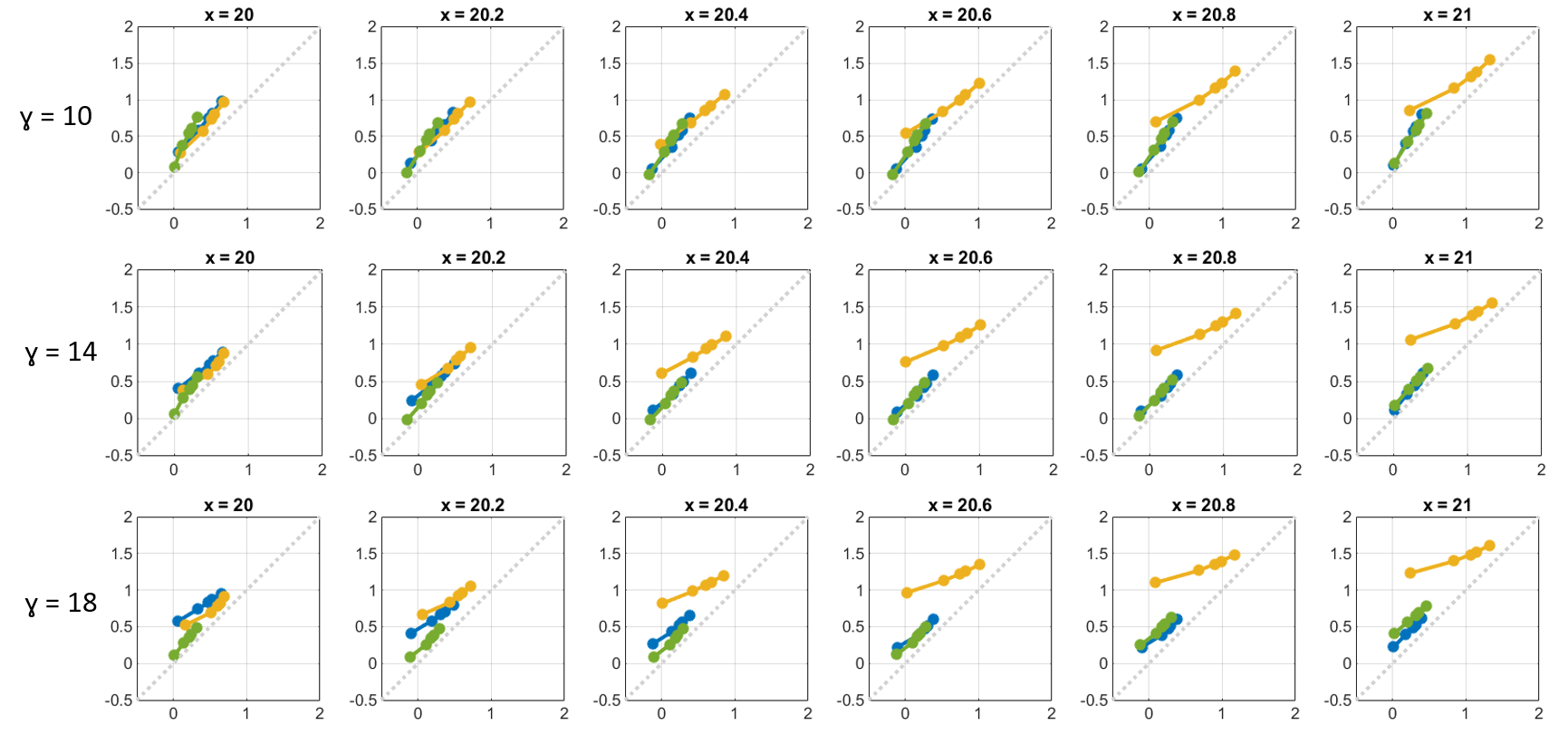}}
\caption{Computations of the inequality that we have proved in Theorem \ref{thm11} are shown for the temperature system. In each plot, the horizontal axis provides estimates of $\text{CVaR}_{\alpha,x}^{\pi_\gamma}(\max_{t=0,1,\dots,T} g_K(X_t))$, and the vertical axis provides estimates of $\textstyle\frac{1}{\gamma}\log\hspace{-0.5mm}\left(\textstyle\frac{1}{\alpha} E_x^{\pi_\gamma}\Big(\sum_{t=0}^T e^{\gamma g_K(X_t)} \Big)\right)$ for 5 different risk-sensitivity levels $\alpha \in \Lambda \coloneqq \{ 0.99, 0.05, 0.01, 0.005, 0.001 \}$. Points associated with smaller values of $\alpha$ (more risk-averse) are positioned farther away from the origin. For a fixed $\gamma$, $\pi_\gamma$ is an optimal (deterministic, Markov) policy for the MDP problem \eqref{Jalphastar}. In each plot, there are three solid colored lines, one for each distribution of the disturbance process (green = no skew, yellow = left skew, blue = right skew). In each plot, $\gamma \in \{10, 14,18\}$ and an initial state $x \in \{20, 20.2,\dots,21\}$ are fixed. The value of $\gamma$ varies along the rows, and the value of $x$ varies along the columns. A dotted gray line of slope 1 is shown for visual comparison.}
\label{linegraphsmanym}
\end{figure*}

Fig. \ref{linegraphsmanym} provides a visual summary of the inequality that we have proved in Theorem \ref{thm11}:
\begin{equation}\label{boundsexample}
    \text{CVaR}_{\alpha,x}^{\pi_\gamma}\hspace{-0.5mm}\Big(\hspace{-0.5mm}\max_{t=0,1,\dots,T} g_K(X_t)\hspace{-0.5mm}\Big) \hspace{-1mm} \leq \hspace{-1mm} \textstyle\frac{1}{\gamma}\log\hspace{-0.5mm}\left(\hspace{-0.5mm}\textstyle\frac{1}{\alpha} E_x^{\pi_\gamma}\hspace{-0.5mm}\Big(\hspace{-0.5mm}\sum_{t=0}^T e^{\gamma g_K(X_t)} \Big) \hspace{-0.5mm}\right)\hspace{-0.5mm}.
\end{equation}
Each plot in Fig. \ref{linegraphsmanym} shows estimates of the right-hand-side of \eqref{boundsexample} on the vertical axis versus estimates of the left-hand-side of \eqref{boundsexample} on the horizontal axis for the 5 values of $\alpha$ in $\Lambda$. In each plot, each solid colored line consists of 5 points, one for each $\alpha \in \Lambda$. Points associated with smaller values of $\alpha$ (more risk-averse) are positioned farther away from the origin. In each plot, there are three solid colored lines, one for each distribution of the disturbance process. In each plot, $\gamma \in \Gamma$ and an initial state $x \in \hat{S}$ are fixed. 
We have chosen initial states inside or on the boundary of the constraint set $K= [20, 21]$ $^{\circ}\text{C}$. Fig. \ref{linegraphsmanym} is consistent with the inequality that we have proved in Theorem \ref{thm11} since the solid colored lines are located above the gray line of slope 1. Fig. \ref{linegraphsmanym} suggests that there is no unique value of $\gamma$ that provides the best approximation for all initial states $x$, risk-sensitivity levels $\alpha$, and disturbance distributions.
%

\begin{figure*}[h]
\centerline{\includegraphics[width=\textwidth]{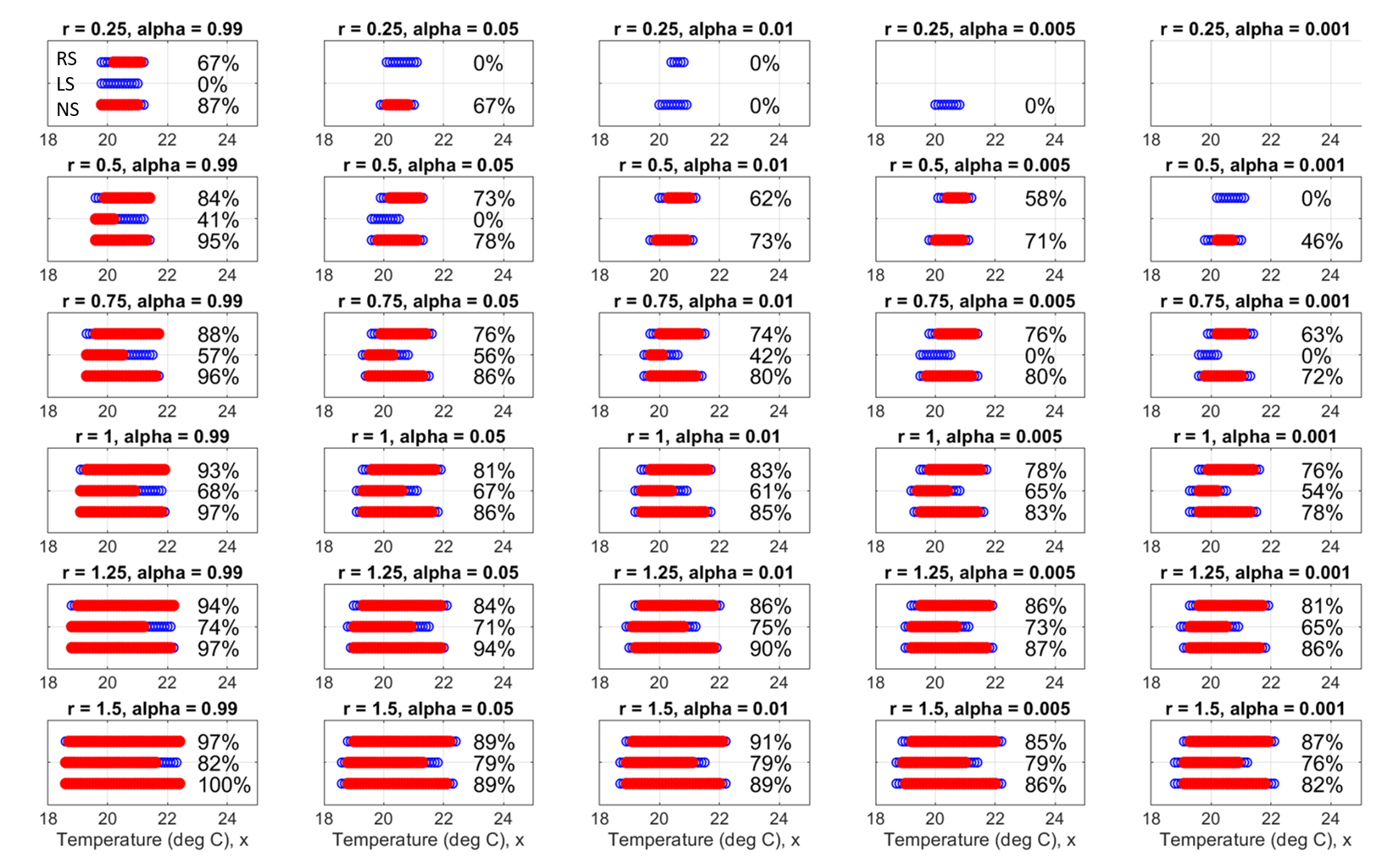}}
\caption{For the temperature system with $\gamma = 14$, estimates of the $(\alpha,r,\gamma)$-under-approximation set $\mathcal{U}_{\alpha,\gamma}^{r} = \mathcal{U}_{\alpha,\gamma}^{r,\pi_\gamma}$ are shown (solid red circles). Estimates of the $(\alpha,r)$-risk-sensitive safe set for the control policy $\pi_\gamma \in \Pi$, $\mathcal{S}_{\alpha}^{r,\pi_\gamma}$, are shown (white circles with blue boundary). Each plot presents the estimated sets for the different disturbance distributions (top interval: right skew (RS), middle interval: left skew (LS), and bottom interval: no skew (NS); see the labels in the first plot). Each percentage $\frac{\text{Number of states in estimate of } \mathcal{U}_{\alpha,\gamma}^{r}}{\text{Number of states in estimate of } \mathcal{S}_{\alpha}^{r,\pi_\gamma}} \cdot 100\%$ indicates the estimated quality of the under-approximation. These percentages are shown whenever the estimate of $\mathcal{S}_{\alpha}^{r,\pi_\gamma}$ is not empty. The risk-sensitivity level $\alpha$ varies from nearly risk-neutral ($\alpha = 0.99$, left-most column) to more risk-averse ($\alpha = 0.001$, right-most column).}
\label{templevelsets}
\end{figure*}

However, by Theorem \ref{thm22}, we have flexibility in choosing the value of $\gamma$. In particular, we favor the quality of the approximations for small values of $\alpha$ due to our focus on safety and present sets using $\gamma = 14$ as an example of a value that reflects this preference (Fig. \ref{templevelsets}).\footnote{Higher-quality approximations are those in which the estimates of the under-approximations are generally closer to the estimates of the risk-sensitive safe sets when considering all three disturbance distributions. We suggest an approach to quantify the quality of the approximations in Fig. \ref{templevelsets}.} Fig. \ref{templevelsets} provides estimates of the $(\alpha,r)$-risk-sensitive safe set for $\pi_\gamma \in \Pi$ \eqref{mySgivenpolicy}
\begin{equation*}\begin{aligned}
\mathcal{S}_{\alpha}^{r,\pi_\gamma} & \coloneqq \left\{x \in S : \text{CVaR}_{\alpha,x}^{\pi_\gamma}\Big(\max_{t = 0,1,\dots,T}  g_K(X_t)\Big) \leq r \right\}
\end{aligned}\end{equation*}
and the $(\alpha,r,\gamma)$-under-approximation set \eqref{myU}
\begin{equation*}\begin{aligned}
\mathcal{U}_{\alpha,\gamma}^{r} 
& = \left\{x \in S : \textstyle\frac{1}{\alpha}E_{x}^{\pi_\gamma}\Big(\sum_{t=0}^T e^{\gamma g_K(X_t)}\Big) \leq e^{\gamma r}\right\}\\
&= \left\{x \in S : {\textstyle\frac{1}{\gamma}}\log\Big(\textstyle\frac{1}{\alpha}E_{x}^{\pi_\gamma}\Big(\sum_{t=0}^T e^{\gamma g_K(X_t)}\Big)\Big) \leq r\right\}.
\end{aligned}\end{equation*}
Note that 
$\mathcal{U}_{\alpha,\gamma}^{r} = \mathcal{U}_{\alpha,\gamma}^{r,\pi_\gamma}$.\footnote{Recall that $\pi_\gamma \in \Pi$ is a policy that satisfies $J_\gamma^*(x) = J_\gamma(x,\pi_\gamma) \; \forall x \in S$. That is, $\pi_\gamma$ is an optimal policy for the MDP problem that defines $\mathcal{U}_{\alpha,\gamma}^{r}$.} In Fig. \ref{templevelsets}, estimates of $\mathcal{U}_{\alpha,\gamma}^{r, \pi_\gamma}$ (solid red) and $\mathcal{S}_{\alpha}^{r,\pi_\gamma}$ (white circles with blue boundary) are shown for the risk-sensitivity levels $\alpha \in \Lambda$ and various $r \in \mathbb{R}$ with $\gamma = 14$. The estimates of $\mathcal{U}_{\alpha,\gamma}^{r,\pi_\gamma}$ are subsets of the estimates of $\mathcal{S}_{\alpha}^{r,\pi_\gamma}$, which we expect by Theorem \ref{thm22}. The estimates of $\mathcal{S}_{\alpha}^{r,\pi_\gamma}$ form an increasing sequence of subsets as $\alpha$ increases and $r$ increases, which is consistent with Lemma \ref{mylemma2}.
%
\subsection{Stormwater System}\label{watersysex}
Next, we illustrate risk-sensitive safety analysis using a gravity-driven stormwater system with an automated valve. 
Consider a two-tank stormwater system evolving on a finite-time horizon $t = 0,1,\dots,T-1$ using a deterministic Markov policy $\pi = (\pi_0,\pi_1,\dots,\pi_{T-1})$, $X_{t+1} = X_t + \bar{f}(X_t,\pi_t(X_t),D_t) \cdot \triangle\tau$. Let $\mathbb{R}^n_+ \coloneqq \{y = (y_1,\dots,y_n)^\text{T} \in \mathbb{R}^n : y_i \geq 0 \; \forall i \}$. The state $X_t$ is the $\mathbb{R}_{+}^2$-valued random water elevations in the tanks at time $t$ (ft, ft). $\pi_t(X_t)$ is the $[0,1]$-valued valve setting at time $t$ (closed to open). $(D_0,D_1,\dots,D_{T-1})$ is a $\mathbb{R}_{+}$-valued, iid stochastic process of surface runoff. $\triangle \tau$ is the duration of $[t,t+1)$. The function $\bar{f} : \mathbb{R}_{+}^2 \times [0,1] \times \mathbb{R}_{+} \rightarrow \mathbb{R}_{+}^2$ is given by
\begin{equation*}\begin{aligned}
\bar{f}(x,u,d) & \coloneqq \left[ \frac{d - q_\text{valve}(x,u)}{a_1}, \frac{d+q_\text{valve}(x,u)-q_\text{drain}(x)}{a_2} \right]^{\text{T}} \\
q_\text{valve}(x,u) & \coloneqq u \cdot \bar{\pi} r_\text{v}^2 \cdot \sgn{h(x)} \cdot \sqrt{2 \bar{g} |h(x)|}  \\
h(x) & \coloneqq \max(x_1 - z_1,0) - \max(x_2 - z_{1,\text{in}},0)\\
q_\text{drain}(x) & \coloneqq \begin{cases} c_\text{d} \bar{\pi} r_\text{d}^2 \sqrt{2 \bar{g} (x_2 - z_2)} & \text{if } x_2 \geq z_2 \\
0 & \text{otherwise}. \end{cases}
\end{aligned}\end{equation*}
Model parameters are in Table \ref{tanksysinfo}. The constraint set $K= [0, k_1] \times [0, k_2]$ specifies the maximum water elevations that the tanks can hold without surcharge. The stage cost $g_K(x) = \max(x_{1} - k_1, x_{2} - k_2, 0)$ is the maximum surcharged water level when the system occupies the state $x \in \mathbb{R}_{+}^2$. 
%

We have identified a discrete distribution for the disturbance process with the approximate statistics, mean (12.2 cfs), variance (9.9 cfs$^2$), and skew (0.74), where cfs is cubic feet per second. In previous work, we obtained runoff samples by simulating a design storm in PCSWMM (Computational Hydraulics International), which extends the US Environmental Protection Agency's Stormwater Management Model \cite{sustech, swmm}. In this previous work, the empirical distribution had positive skew, and the mean was about 12.2 cfs \cite{sustech}, which are reflected in the current distribution (not shown in the interest of space).

\begin{figure*}[h]
\centerline{\includegraphics[width=\textwidth]{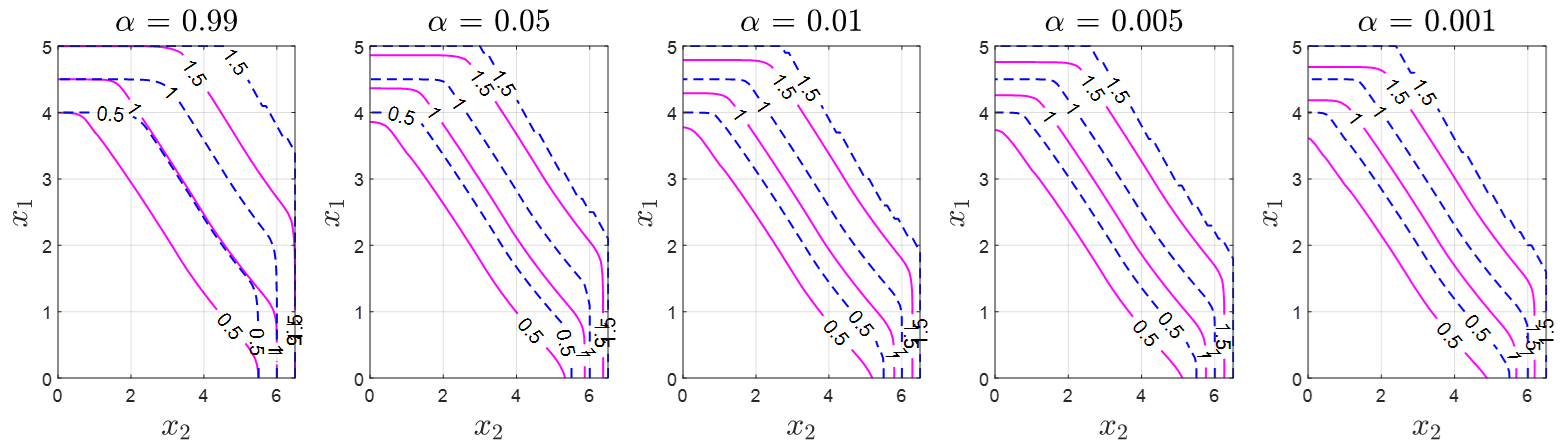}}
\caption{For the stormwater system with $\gamma = 22$, estimates of the boundary of the $(\alpha,r,\gamma)$-under-approximation set $\mathcal{U}_{\alpha,\gamma}^{r} = \mathcal{U}_{\alpha,\gamma}^{r,\pi_\gamma}$ are shown (solid pink). Estimates of the boundary of the $(\alpha,r)$-risk-sensitive safe set for the control policy $\pi_\gamma \in \Pi$, $\mathcal{S}_{\alpha}^{r,\pi_\gamma}$, are shown (dotted blue). We present $r \in \{0.5, 1, 1.5\}$ and $\alpha \in \{0.99, 0.05, 0.01, 0.005, 0.001\}$. The percentages $\frac{\text{Number of states in estimate of } \mathcal{U}_{\alpha,\gamma}^{r}}{\text{Number of states in estimate of } \mathcal{S}_{\alpha}^{r,\pi_\gamma}} \cdot 100\%$ indicate the estimated quality of the under-approximations. We list these percentages for the plots in this figure in Table \ref{tablewaterpercent}.}
\label{waterlevelfig}
\end{figure*}

In Fig. \ref{waterlevelfig}, we show estimates of risk-sensitive safe sets and their under-approximations using $\gamma = 22$ for 5 risk-sensitivity levels (see also Table \ref{tablewaterpercent}). The shape of the contour of $\mathcal{S}_{\alpha}^{r,\pi_\gamma}$ indicates a critical trade-off between the maximum initial water elevations in the two tanks from which the system meets a desired degree of safety. The similarity in the shapes of $\mathcal{S}_{\alpha}^{r,\pi_\gamma}$ and $\mathcal{U}_{\alpha,\gamma}^{r}$ is notable, suggesting that $\mathcal{U}_{\alpha,\gamma}^{r}$ may be a useful tool for inferring these critical trade-offs in networked water systems. 
%
%
\begin{table}[h]
\caption{Stormwater System Parameters}
\label{table}
\setlength{\tabcolsep}{3pt}
\begin{tabular}{|p{25pt}|p{125pt}|p{65pt}|}
\hline
Symbol &  Description & Value \\
\hline
$a_1$ & surface area of tank 1 & 28292 ft\textsuperscript{2} \\
$a_2$ & surface area of tank 2 & 25965 ft\textsuperscript{2} \\
$c_d$ & discharge coefficient & 0.61 (no units) \\
$\bar{g}$ & acceleration due to gravity & 32.2 $\frac{\text{ft}}{\text{s}^2}$\\
$k_1$ & maximum water level in tank 1 & 3.5 ft \\
$k_2$ & maximum water level in tank 2 & 5 ft \\
$\bar{\pi}$ & circle circumference-to-diameter ratio & $\approx$ 3.14 \\
$r_\text{d}$ & radius of drain & $2/3$ ft \\
$r_\text{v}$ & radius of valve & $1/3$ ft\\
$ \triangle \tau $ & duration of $[t,t+1)$ & 5 min \\
$T$ & length of discrete time horizon & 24 ($=$ 2 h) \\
$A$ & control space & $[0,1]$ (no units)\\ 
$S$ & state space & $[0, 5]\text{ ft}\times [0,6.5]\text{ ft}$ \\
$z_1$ & invert elevation of pipe from base of tank 1 & 1 ft\\
$z_{1,\text{in}}$ & invert elevation of pipe from base of tank 2 & 2.5 ft\\
$z_2$ & elevation from base of tank 2 to orifice & 1 ft\\
\hline
\multicolumn{3}{p{251pt}}{ft $=$ feet, s $=$ 
seconds, min $=$ minutes, h $=$ hours.}
\end{tabular}
\label{tanksysinfo}
\end{table}
%
%
\newpage\section{Concluding Remarks}\label{conclusion}
This paper develops trajectory-wise safety specifications for control systems that quantify \textcolor{black}{the severity of random} harmful outcomes and thereby generalize classical stochastic safety analysis.
Our primary contribution is to develop a tractable, interpretable safety analysis method with theoretical guarantees that assesses the upper tail of a cost distribution by using CVaR. It is notable that our method provides a parameter-dependent upper bound to the CVaR of a maximum cost without augmenting the state space. We have developed compelling numerical examples, which demonstrate the utility and tractability of our under-approximation approach. 
Moreover, we have proposed a risk-sensitive safe set definition in terms of a new coherent risk functional, inspired by CVaR, that admits a parameter-independent upper bound. We show that this upper bound can be computed via DP on the original state space by proving the regularity of a supremum over a function space for a class of \textcolor{black}{transition kernels}. Numerical investigations of leveraging our approximation to provide an efficient preliminary estimate to the exact CVaR is an exciting future direction. For instance, we have recently demonstrated the usefulness of efficient approximate ``warm-start'' computations to examine the effect of different design changes to stormwater infrastructure \cite{chapmantac2022}. More broadly, combining techniques from approximate dynamic programming, stochastic rollout, and risk-sensitive safety analysis could lead to novel controller synthesis algorithms for higher-dimensional systems.
%
\begin{table}[h]
\label{table}
\caption{}
 \setlength{\tabcolsep}{3pt}
\begin{tabular}{|p{35pt}|p{35pt}|p{35pt}|p{35pt}|p{38pt}|p{37pt}|}
\hline
\textcolor{white}{$r = 0.5$} & $\alpha = 0.99$ & $\alpha = 0.05$ & $\alpha = 0.01$ & $\alpha = 0.005$ & $\alpha = 0.001$ \\
\hline
$r = 0.5$ ft & 74.3 $\%$ & 77.3 \%  & 76.6 \%  & 76.0  \% & 72.5 \%  \\
\hline
$r = 1$ ft & 82.9 \% & 84.2 \%  & 83.1 \%  & 83.0 \%  & 81.2 \%  \\
\hline
$r = 1.5$ ft & 84.4 \% & 75.6 \%  & 71.2 \%  & 69.9 \%  & 66.0 \%  \\
\hline
\multicolumn{6}{p{240pt}}{This table provides the percentages $\frac{\text{Number of states in estimate of } \mathcal{U}_{\alpha,\gamma}^{r}}{\text{Number of states in estimate of } \mathcal{S}_{\alpha}^{r,\pi_\gamma}} \cdot 100\%$ for the sets in Fig. \ref{waterlevelfig} (stormwater system, $\gamma = 22$).}
\end{tabular}
\label{tablewaterpercent}
\end{table}
\section*{Appendix}
\begin{lemma}[Attainment of Supremum]\label{attainmentlemma}
Let $J : S \rightarrow \mathbb{R}$ be Borel measurable and bounded, and let $\alpha \in (0,1]$. Define the function $\phi : S \times A \rightarrow \mathbb{R}$ by
\begin{equation}\label{myphiphi}
    \phi(x,u) \coloneqq  \sup \left\{ \int_{S} J \xi \; \mathrm{d}Q( \cdot|x,u) : \xi \in \mathcal{R}_{\alpha}(x,u)\right\}.
\end{equation}
Then, for any $(x,u) \in S \times A$, there is a $\xi^*(\cdot|x,u) \in \mathcal{R}_\alpha(x,u)$ such that
\begin{equation}
    \phi(x,u) =  \int_{S}   J \xi^*(\cdot|x,u) \; \mathrm{d} Q(\cdot|x,u).\vspace{1mm}
\end{equation}
\end{lemma}
\begin{proof}
Let $(x,u) \in S \times A$, and fix the probability space $(S, \mathcal{B}(S), Q(\cdot|x,u))$. Denote $L^p_{x,u} \coloneqq L^p(S, \mathcal{B}(S), Q(\cdot|x,u))$ for brevity, and view $\mathcal{R}_\alpha(x,u)$ as a subset of $L^2_{x,u}$ \textcolor{black}{with the weak topology}. Define the functional $\psi : L^2_{x,u} \rightarrow \mathbb{R}$ by
\begin{equation} \psi(\xi) \coloneqq \int_{S} J \xi \; \mathrm{d}Q(\cdot|x,u).
\end{equation} 
It suffices to show that $\psi$ is weakly continuous and $\mathcal{R}_\alpha(x,u)$ is weakly compact. \textcolor{black}{Weak continuity follows from two well-known facts: 1) a linear functional on a normed vector space is weakly continuous if and only if it is strongly continuous \cite[Prop. 2.5.3]{Megginson}, and 2) a linear functional on a normed vector
space is strongly continuous if and only if it is bounded \cite[Prop. 5.2]{folland2013real}.} By applying standard techniques, it follows that $\psi$ is a bounded linear functional on a normed vector space, and thus, $\psi$ is weakly continuous. \textcolor{black}{As $\mathcal{R}_\alpha(x,u)$ is a bounded and weakly closed subset of $L^2_{x,u}$, $\mathcal{R}_\alpha(x,u)$ is weakly compact by the Banach-Alaoglu Theorem \cite[p. 401]{shapiro2009lectures}. Here, we use the fact that $L^2_{x,u}$ is reflexive, and hence, the weak and weak* topologies of $L^2_{x,u}$ are the same. We provide details about weak closedness in a footnote.}\footnote{For weak closedness, recall the fact \cite[Thm. 3.7]{brezis2010}: \emph{Let $E$ be a Banach space, and let $C$ be a convex subset of $E$. Then, $C$ is closed in the weak topology if and only if it is closed in the strong topology.} Since $\mathcal{R}_\alpha(x,u) \subseteq L^2_{x,u}$ is convex, to show that $\mathcal{R}_\alpha(x,u)$ is weakly closed, it suffices to show that $\mathcal{R}_\alpha(x,u)$ is strongly closed. \textcolor{black}{Strong closedness of $\mathcal{R}_\alpha(x,u)$ follows from 1) strong convergence implying weak convergence and 2) strong convergence implying the existence of a subsequence that converges a.e. to the same limit function \cite[Thms. 2.5.1 \& 2.5.3]{ash1972}. Let $(\xi_n)_{n \in \mathbb{N}} \in \mathcal{R}_\alpha(x,u)$ converge strongly to $\xi^* \in L^2_{x,u}$. The first fact ensures that $\int_S \xi^* \mathrm{d}Q(\cdot|x,u) = 1$, and the second fact ensures that $0 \leq \xi^* \leq \alpha^{-1/T}$ a.e., and thus, $\xi^* \in \mathcal{R}_\alpha(x,u)$.}\label{myfootnotenote}} 
\end{proof}

We use similar techniques to prove Lemma \ref{existweak}. Lemma \ref{existweak} is needed to guarantee that a supremum over $\mathcal{R}_\alpha(x,u)$ \eqref{myphiphi} is upper semi-continuous in $(x,u)$.
\begin{lemma}[Existence of weakly convergent subsequence]\label{existweak}
Let $\mu$ be a probability measure on $(S, \mathcal{B}(S))$,  $\mathcal{G}_\alpha(\mu)$ the set of functions $\xi \in L^2_{\mu} \coloneqq L^2(S, \mathcal{B}(S), \mu)$ such that $\xi \in [0, \alpha^{-1/T}]$ a.e. w.r.t. $\mu$, and $(\xi_n)_{n \in \mathbb{N}} \subseteq \mathcal{G}_\alpha(\mu)$. Then, there exist $(\xi_{n_k})_{k \in \mathbb{N}} \subseteq (\xi_n)_{n \in \mathbb{N}}$ and $\xi^* \in \mathcal{G}_\alpha(\mu)$ such that $(\xi_{n_k})_{k \in \mathbb{N}}$ converges to $\xi^*$ in the weak topology of $L^2_{\mu}$.
\end{lemma}
\begin{proof}
The proof requires two facts. The first fact is \cite[Thm. 3.18]{brezis2010}: \emph{Assume that $E$ is a reflexive Banach space, and let $(x_n)$ be a (uniformly) bounded sequence in $E$. Then, there is a subsequence $(x_{n_k})\subseteq (x_{n})$ that converges in the weak topology.} $L^2_\mu$ is a reflexive Banach space, and $||\xi_n||_{L^2_\mu} \leq \alpha^{-1/T}$ for all $n \in \mathbb{N}$. Thus, there exist $(\xi_{n_k})_{k \in \mathbb{N}} \subseteq (\xi_n)_{n \in \mathbb{N}}$ and $\xi^* \in L^2_\mu$ such that $(\xi_{n_k})_{k \in \mathbb{N}}$ converges weakly to $\xi^*$.
Moreover, it holds that $\xi^* \in \mathcal{G}_\alpha(\mu)$ using \cite[Thm. 3.7]{brezis2010} (Footnote \ref{myfootnotenote}). Indeed, $\mathcal{G}_\alpha(\mu)$ is a convex subset of $L^2_\mu$, and $\mathcal{G}_\alpha(\mu)$ is strongly closed in $L^2_\mu$. 
Thus, $\mathcal{G}_\alpha(\mu)$ is weakly closed in $L^2_\mu$, which implies that $\xi^* \in \mathcal{G}_\alpha(\mu)$.
\end{proof}

We use Lemma \ref{existweak} to prove the next supporting result.
\begin{lemma}[Properties of $\phi$]\label{keypropofphi}
Let $J : S \rightarrow \mathbb{R}$ be Borel measurable and bounded and $\alpha \in (0,1]$. Under Assumption \ref{assumption3}, $\phi$ \eqref{myphiphi} is upper semi-continuous (usc) and bounded.
\end{lemma}
\begin{proof}
Boundedness of $\phi$ follows from $Q(\cdot|x,u)$-a.e.-boundedness of $J \xi$ for any $\xi \in \mathcal{R}_{\alpha}(x,u)$. Now, $\phi$ is usc if and only if
\begin{equation*}
\mathcal{C}_a \coloneqq \big\{(x,u) \in S \times A : \phi(x,u) \geq a \big\}
\end{equation*}
is closed for every $a \in \mathbb{R}$. Let $a \in \mathbb{R}$ and $(x_n,u_n)_{n \in \mathbb{N}} \subseteq \mathcal{C}_a$ converging to $(x,u) \in S \times A$ be given, and we shall show that $(x,u) \in \mathcal{C}_a$. It suffices to show that there exist $(x_{n_k},u_{n_k})_{k \in \mathbb{N}} \subseteq (x_n,u_n)_{n \in \mathbb{N}}$ and $(c_k)_{k \in \mathbb{N}} \subseteq \mathbb{R}$ with $c_k \rightarrow 0$ such that 
\begin{equation*} 
\phi(x_{n_k},u_{n_k}) \leq c_k + \phi(x,u) \;\;\; \forall k \in \mathbb{N}. 
\end{equation*}
Indeed, if so, then 
\begin{equation*}
a \leq  \limsup_{k \rightarrow \infty} \phi(x_{n_k},u_{n_k}) \leq \limsup_{k \rightarrow \infty} c_k + \phi(x,u) = \phi(x,u). 
\end{equation*}
Denote $z_n \coloneqq (x_n,u_n)$ and $z \coloneqq (x,u)$ for brevity. By Lemma \ref{attainmentlemma}, for every $n \in \mathbb{N}$, 
\begin{equation*}
  \exists\xi_n \coloneqq \xi^*(\cdot|z_n) \in \mathcal{R}_\alpha(z_n) \text{ s.t. } \phi(z_n) = \int_S J \xi_n \;\mathrm{d}Q(\cdot|z_n).  
\end{equation*} 
Since $\xi_n \in [0,\alpha^{-1/T}]$ a.e. w.r.t. $Q(\cdot|z_n)$,
\begin{equation*}
\exists B(z_n) \in \mathcal{B}(S) \; \text{ s.t. } \; \xi_n(y) \in [0,\alpha^{-1/T}] \;\;\; \forall y \in B(z_n),     
\end{equation*}
where $Q(S\setminus B(z_n)|z_n) = 0$. Define
\begin{equation*}
    \tilde \xi_n \coloneqq I_{B(z_n)} \xi_n.
\end{equation*} 
It follows that $\tilde \xi_n \in \mathcal{R}_\alpha(z_n)$ with $\tilde \xi_n \in [0,\alpha^{-1/T}]$ everywhere. Also, it holds that $(\tilde \xi_n)_{n \in \mathbb{N}} \subseteq \mathcal{G}_\alpha(Q(\cdot|z))$, where
\begin{equation*}
  \mathcal{G}_\alpha(Q(\cdot|z))  \coloneqq \big\{ \xi \in L^2_z : \xi \in [0,\alpha^{-1/T}] \text{ a.e. w.r.t. }Q(\cdot|z) \big\}
\end{equation*}
and $L^2_z \coloneqq L^2(S, \mathcal{B}(S),Q(\cdot|z))$. By Lemma \ref{existweak}, there exist $(\tilde \xi_{n_k})_{k \in \mathbb{N}} \subseteq (\tilde \xi_n)_{n \in \mathbb{N}}$ and $\xi^\dagger \in \mathcal{G}_\alpha(Q(\cdot|z))$ such that $(\tilde \xi_{n_k})_{k \in \mathbb{N}}$ converges to $\xi^\dagger$ in the weak topology of $L^2_z$. 
It holds that $\xi^\dagger \in \mathcal{R}_\alpha(z)$, and we explain why $\int_S \xi^\dagger  \mathrm{d}Q(\cdot|z) = 1$ next. For any $k \in \mathbb{N}$, it holds that $|\tilde \xi_{n_k}| \leq \alpha^{-1/T}$ everywhere, and it follows that
\begin{equation*}
   \left | \int_S \xi^\dagger \mathrm{d}Q(\cdot|z) - 1 \right| \leq \text{Term1}(k) + \text{Term2}(k),
\end{equation*}
where
\begin{equation*}\begin{aligned}
    \text{Term1}(k) & \coloneqq \left| \int_S \xi^\dagger \mathrm{d}Q(\cdot|z) - \int_S \tilde \xi_{n_k} \mathrm{d}Q(\cdot|z) \right| \\
    \text{Term2}(k) & \coloneqq \alpha^{-1/T} \big|Q(\cdot|z) - Q(\cdot|z_{n_k})\big|(S).
\end{aligned}\end{equation*}
The quantity $|Q(\cdot|z) - Q(\cdot|z_{n_k})|(S)$ is the total variation of the signed measure $Q(\cdot|z) - Q(\cdot|z_{n_k})$ evaluated at the set $S$. By the weak convergence of $(\tilde \xi_{n_k})_{k \in \mathbb{N}}$ to $\xi^\dagger$ in $L^2_z$, $\text{Term1}(k) \rightarrow 0$ as $k \rightarrow \infty$. By Assumption \ref{assumption3}, $\text{Term2}(k) \rightarrow 0$ as $k \rightarrow \infty$. 

Now, for every $k \in \mathbb{N}$, we use the triangle inequality and everywhere boundedness of $J \tilde \xi_{n_k}$ to find that 
\begin{equation*}
    \phi(z_{n_{k}}) - \phi(z) \leq \text{Term3}(k) + \text{Term4}(k),
\end{equation*}
where 
\begin{equation*}
    \text{Term3}(k) \coloneqq \frac{b}{\alpha^{1/T}} \big|Q(\cdot|z_{n_{k}})-Q(\cdot|z) \big|(S),
\end{equation*}
$b \in \mathbb{R}$ satisfies $|J(y)|\leq b$ for all $y \in S$, and
\begin{equation*}
\text{Term4}(k) \coloneqq \left |\int_S J \tilde \xi_{n_{k}} \mathrm{d}Q(\cdot|z) - \int_{S} J \xi^\dagger \mathrm{d}Q(\cdot|z) \right|. 
\end{equation*}
By the weak convergence of $(\tilde \xi_{n_k})_{k \in \mathbb{N}}$ to $\xi^\dagger$ in $L^2_z$, $\text{Term4}(k) \rightarrow 0$ as $k \rightarrow \infty$, and by Assumption \ref{assumption3}, $\text{Term3}(k) \rightarrow 0$ as $k \rightarrow \infty$. We choose 
\begin{equation*}
    c_k \coloneqq \text{Term3}(k) + \text{Term4}(k) \;\;\; \forall k \in \mathbb{N},
\end{equation*} 
and it follows that $\phi$ is usc.
\end{proof}

We use the upper semi-continuity of $\phi$ to prove Theorem \ref{thm33}.

\begin{proof}[Theorem \ref{thm33}]
Proceed by induction. $J_T^\alpha = c_T$ is usc and bounded. Now, assume that for some $t = T-1,\dots,1,0$, $J_{t+1}^\alpha$ is usc and bounded. Then, $J_{t+1}^\alpha$ is Borel measurable and bounded, which implies that $\varphi_t^\alpha$ is usc and bounded by Lemma \ref{keypropofphi}. Since $v_{t}^\alpha = c_t + \varphi_t^\alpha$ is a sum of usc and bounded functions, $v_t^\alpha$ is usc and bounded. By \cite[Prop. 7.34]{bertshreve2004}, we conclude that $J_t^\alpha$ is usc and bounded, and for every $\epsilon > 0$, there is a Borel-measurable function $\kappa_t^{\alpha,\epsilon} : S \rightarrow A$ such that $J_{t}^\alpha(x) \leq v_{t}^\alpha(x,\kappa_t^{\alpha,\epsilon}(x)) \leq J_{t}^\alpha(x) + \epsilon$ for all $x \in S$.

\textcolor{black}{A DP argument completes the proof, which we outline below.\footnote{A conditional expectation is not unique everywhere in general \cite[Th. 6.3.3]{ash1972}. However, for the sake of simplicity, we write that a relation with a conditional expectation holds everywhere, following the proof of \cite[Th. 3.2.1]{hernandez2012discrete}.}} Let $\Pi'$ be the set of randomized Markov policies. For $t = 0,1,\dots,T$, define the random cost-to-go by 
\begin{equation*}
  Y_t \coloneqq  \begin{cases} c_T(X_T) + \sum_{i=t}^{T-1} c_i(X_i,U_i) & \text{if }t < T \\ c_T(X_T) & \text{if } t = T, \end{cases}
\end{equation*}
and note that $Y = Y_0$. For any $\pi \in \Pi'$ and $\zeta \in \mathcal{D}_\alpha$, we denote the $(\pi,\zeta)$-conditional expectation of $Y_t$ given $X_t$ by $W_t^{\pi,\zeta}(x_t) \coloneqq E^{\pi,\zeta}(Y_t|X_t = x_t)$, where $x_t \in S$. For any $\pi = (\pi_0,\pi_1,\dots,\pi_{T-1}) \in \Pi'$ and $\zeta = (\xi_0,\xi_1,\dots,\xi_{T-1}) \in \mathcal{D}_\alpha$, the following recursion (``law of iterated expectations'') holds: for $t = 0,1,\dots,T-1$ and $x \in S$,
\begin{subequations}\label{lawiter}
\begin{equation}
    W_t^{\pi,\zeta}(x) =  \int_A \big( c_t(x,u) +  \psi_{t}^{\pi,\zeta}(x,u) \big) \; \pi_t(\mathrm{d}u|x),
\end{equation}
where $\psi_{t}^{\pi,\zeta}$ is defined by
\begin{equation}
    \psi_{t}^{\pi,\zeta}(x,u) \coloneqq  \int_S W_{t+1}^{\pi,\zeta}(y) \; \xi_t(y|x,u) \; Q(\mathrm{d}y|x,u) 
\end{equation}
\end{subequations}
with $\xi_t(\cdot|x,u) \in \mathcal{R}_\alpha(x,u)$ for each $(x,u) \in S \times A$. For any policy $\pi \in \Pi'$, we have
\begin{equation}\label{almostlast}
    \rho_{\alpha,x}^\pi(Y) = \sup_{\zeta \in \mathcal{D}_\alpha} W_0^{\pi,\zeta}(x) \;\;\; \forall x \in S.
\end{equation}
Let $\epsilon > 0$ be given. Then, for each $t = 0,1, \dots, T-1$, there exists a Borel-measurable function $\mu_t^{\alpha,\epsilon} : S \rightarrow A$ such that 
\begin{equation}\label{controllaw}
    J_{t}^\alpha(x) \leq v_{t}^\alpha(x,\mu_t^{\alpha,\epsilon}(x)) \leq J_{t}^\alpha(x) + \frac{\epsilon}{T} \;\;\; \forall x \in S.
\end{equation}
Define $\pi_\epsilon^* \coloneqq (\mu_0^{\alpha,\epsilon}, \dots, \mu_{T-1}^{\alpha,\epsilon}) \in \Pi'$, which is a deterministic Markov policy and thus is an element of $\Pi$ (the class of randomized history-dependent policies) as well. Hence,
\begin{equation*}
\inf_{\pi \in \Pi} \rho_{\alpha,x}^\pi(Y) \leq \rho_{\alpha,x}^{\pi_\epsilon^*}(Y) \overset{\eqref{almostlast}}{=} \sup_{\zeta \in \mathcal{D}_\alpha} W_0^{\pi_\epsilon^*,\zeta}(x) \;\;\; \forall x \in S. 
\end{equation*}
It suffices to prove that 
\begin{equation}\label{mymy38}
    W_t^{\pi_\epsilon^*,\zeta}(x) \leq J_t^\alpha(x) +  \frac{(T-t)\epsilon}{T}
\end{equation}
for all $x \in S$, $\zeta \in \mathcal{D}_\alpha$, and $t \in\{ 0, 1, \dots, T\}$.
Indeed, by setting $t = 0$ in \eqref{mymy38} and taking the supremum over $\mathcal{D}_\alpha$, we would derive $\rho_{\alpha,x}^{\pi_\epsilon^*}(Y) \leq J_0^\alpha(x) + \epsilon$ $\forall x \in S$. Since $\epsilon > 0$ is arbitrary, the desired statement would be shown. \textcolor{black}{The sufficient condition \eqref{mymy38} holds by an inductive argument.}\footnote{The base case holds because $W_T^{\pi_\epsilon^*, \zeta} = c_T = J_T^\alpha$ for all $\zeta \in \mathcal{D}_\alpha$. Now, assume that for some $t \in \{0,1,\dots,T-1\}$, it holds that $W_{t+1}^{\pi_\epsilon^*, \zeta}(x) \leq J_{t+1}^\alpha(x) + \textstyle \frac{(T-t-1)\epsilon}{T}$ for all $x \in S$ and $\zeta \in \mathcal{D}_\alpha$. Let $x \in S$ and $\zeta = (\xi_0,\xi_1,\dots,\xi_{T-1}) \in \mathcal{D}_\alpha$ be given. Since $\pi_\epsilon^*$ is a deterministic Markov policy, we have $$W_t^{\pi_\epsilon^*,\zeta}(x) \overset{\eqref{lawiter}}{=} c_t(x,\mu_t^{\alpha,\epsilon}(x)) +  \psi_{t}^{\pi_\epsilon^*,\zeta}(x,\mu_t^{\alpha,\epsilon}(x)).$$ By the induction hypothesis and $\xi_t(\cdot|x,\mu_t^{\alpha,\epsilon}(x)) \in \mathcal{R}_\alpha(x,\mu_t^{\alpha,\epsilon}(x))$ from the definition of $\mathcal{D}_\alpha$, it follows that
$$\psi_{t}^{\pi_\epsilon^*,\zeta}(x,\mu_t^{\alpha,\epsilon}(x))  \leq \varphi_t^\alpha(x,\mu_t^{\alpha,\epsilon}(x)) + \frac{(T-t-1)\epsilon}{T}.$$ Since $v_{t}^\alpha = c_t + \varphi_t^\alpha$, we derive $ W_t^{\pi_\epsilon^*,\zeta}(x)  \leq \textstyle v_{t}^\alpha(x,\mu_t^{\alpha,\epsilon}(x)) + \frac{(T-t-1)\epsilon}{T}$. Then, we complete the induction using the second inequality in \eqref{controllaw}, namely $v_{t}^\alpha(x,\mu_t^{\alpha,\epsilon}(x)) \leq J_{t}^\alpha(x) + \textstyle \frac{\epsilon}{T}$.}
\end{proof}

\section*{Acknowledgments}
The authors extend their sincerest gratitude to the four anonymous reviewers whose feedback was invaluable in improving the rigor and presentation of this work. The authors gratefully acknowledge many discussions with Jonathan Lacotte, Yuxi Han, Zhiyan Ding, and Michael Lim during the development of an earlier version of this paper. The authors are also thankful for advice provided by Serdar Y\"{u}ksel, Nicole B\"{a}uerle, Ludovic Tangpi, Michael Fau\ss, Alois Pichler, and Raymond Kwong.
\bibliographystyle{ieeecolor}
%

%
%
\vspace{-10mm}
\begin{IEEEbiography}[{\includegraphics[width=1in,height=1.25in,clip,keepaspectratio]{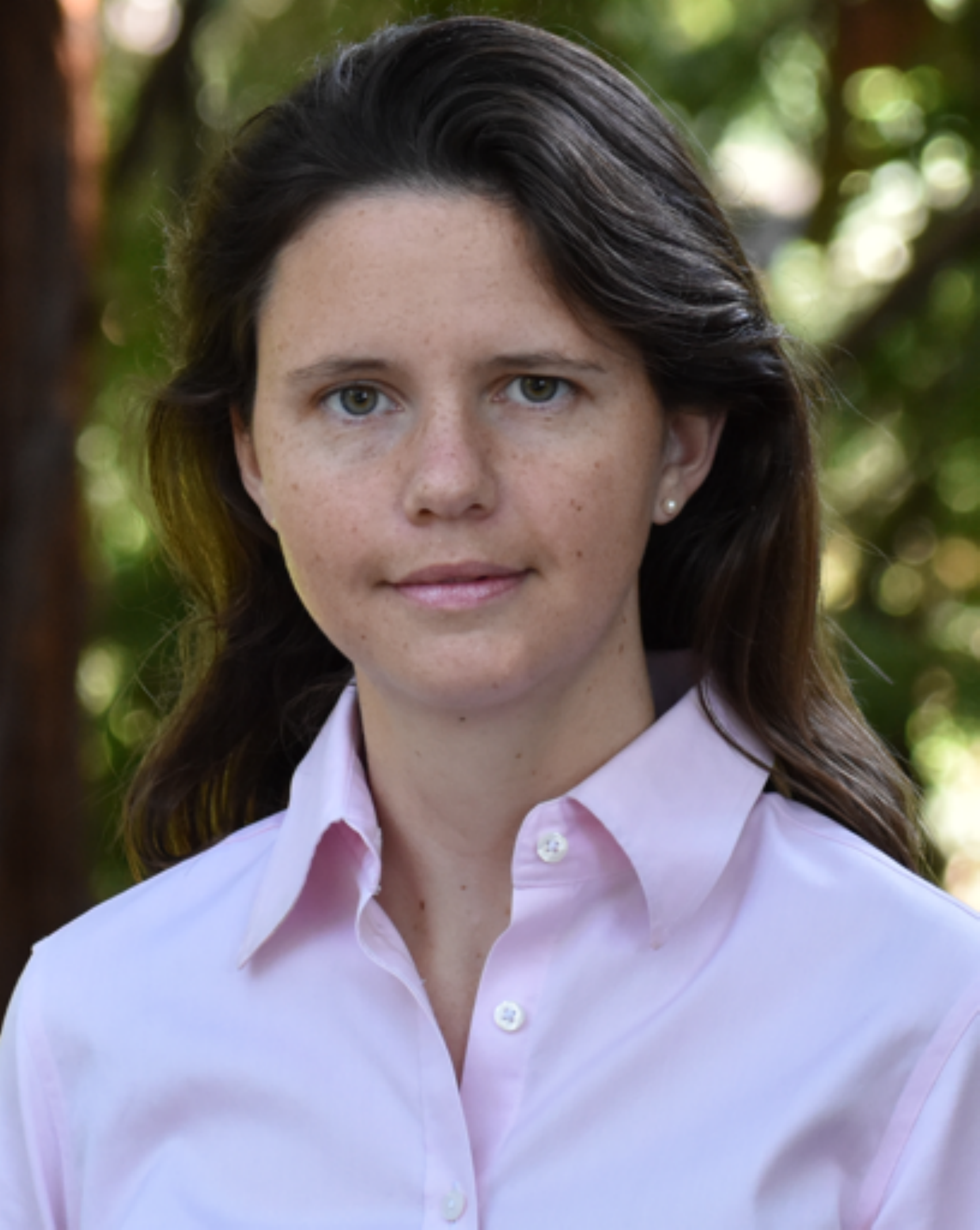}}]{Dr. Margaret Chapman} is an Assistant Professor with the Edward S. Rogers Sr. Department of Electrical and Computer Engineering, University of Toronto, which she joined in July 2020. Her research focuses on risk-averse and stochastic control, with emphasis on safety and applications to healthcare and sustainable cities. Margaret is a recipient of a Leon O. Chua Award for achievement in nonlinear science from UC Berkeley (2021), a US National Science Foundation Graduate Research Fellowship (2014), a Berkeley Fellowship for Graduate Study (2014), and a Stanford University Terman Engineering Scholastic Award (2012). She received her B.S. and M.S. degrees in Mechanical Engineering from Stanford University (2012, 2014) and her Ph.D. degree in EECS from UC Berkeley (2020).\end{IEEEbiography}
\vspace{-5mm}
\begin{IEEEbiography}[{\includegraphics[width=1in,height=1.25in,clip,keepaspectratio]{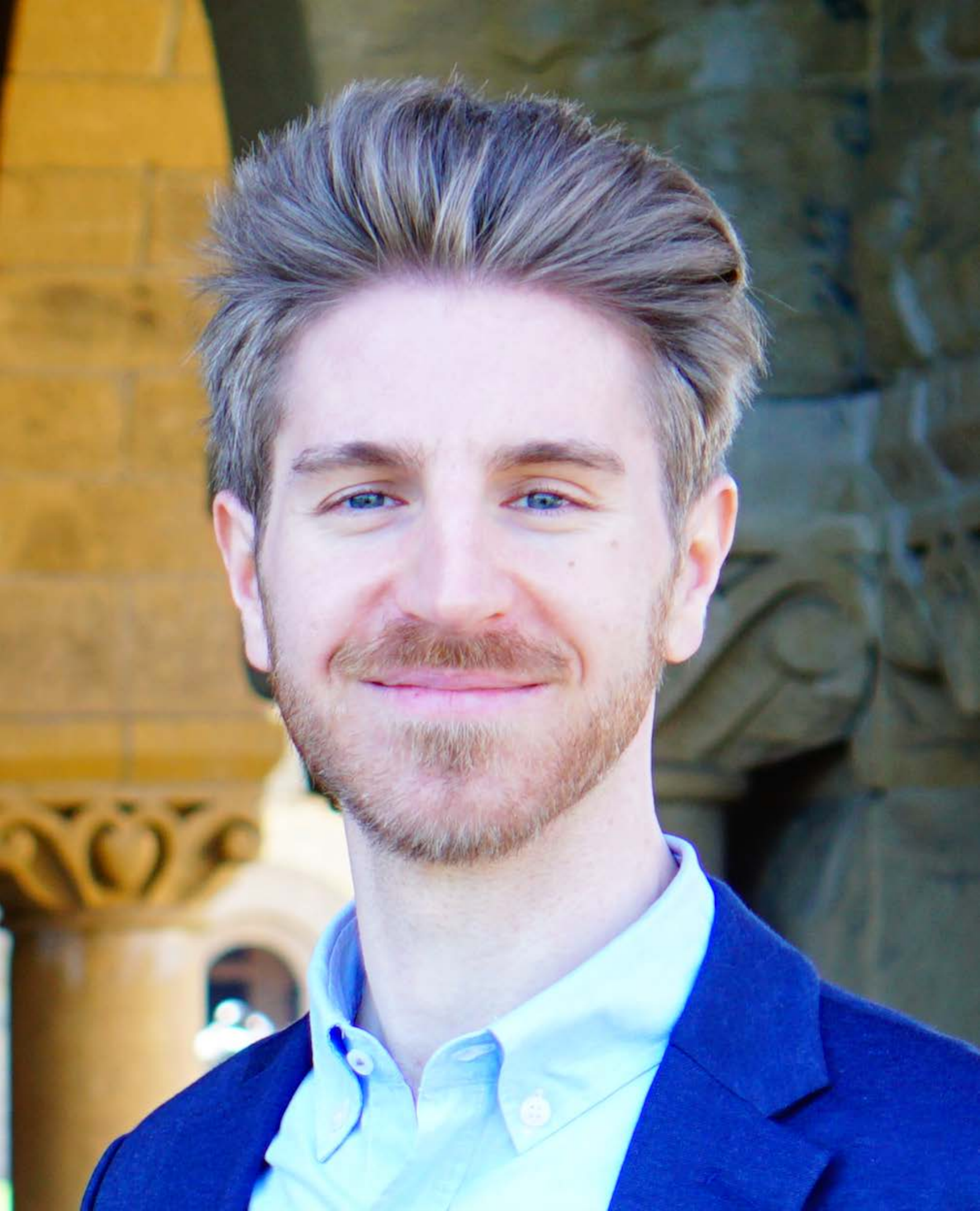}}]{Dr. Riccardo Bonalli} obtained his M.Sc. in Mathematical Engineering from Politecnico di Milano, Italy in 2014 and his Ph.D. in applied mathematics from Sorbonne Universite, France in 2018 in collaboration with ONERA - The French Aerospace Lab, France. He is a recipient of the ONERA DTIS Best Ph.D. Student Award 2018. He was a postdoctoral researcher with the Department of Aeronautics and Astronautics, Stanford University. Currently, Riccardo is a tenured researcher with the Laboratory of Signals and Systems (L2S), Universit\'{e} Paris-Saclay, Centre National de la Recherche Scientifique (CNRS), CentraleSup\'{e}lec, France. His main research interests concern theoretical and numerical robust optimal control with techniques from differential geometry, statistical analysis, and machine learning and applications in aerospace systems and robotics.
\end{IEEEbiography}
\vspace{-5mm}
\begin{IEEEbiography}[{\includegraphics[width=1in,height=1.25in,clip,keepaspectratio]{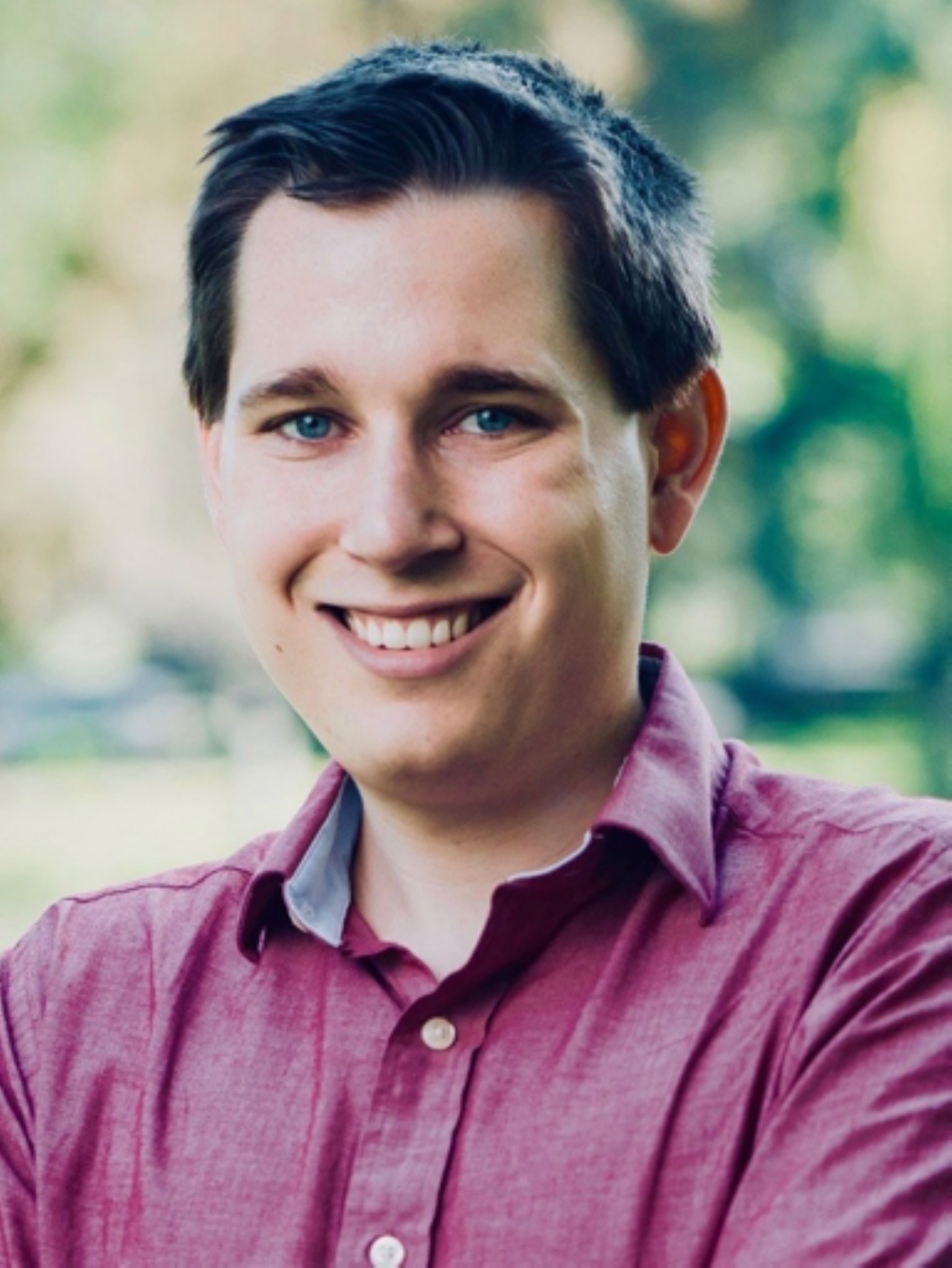}}]{Mr. Kevin Smith} is a Ph.D. candidate in Environmental and Water Resources Engineering at Tufts University and a product developer at OptiRTC, Inc., where he is responsible for developing flexible real-time systems for the continuous monitoring and adaptive control of stormwater infrastructure. Kevin's research seeks to understand the opportunities and risks associated with semi-autonomous civil infrastructure, especially when considered as a technology for mediating environmental conflicts. Kevin is a recipient of the US National Science Foundation Integrative Graduate Education and Research Traineeship (IGERT) on Water and Diplomacy. Kevin earned his B.A. in Environmental Studies from Oberlin College and his B.S. in Earth and Environmental Engineering from Columbia University.\end{IEEEbiography}
\vspace{-5mm}
%
%
\begin{IEEEbiography}[{\includegraphics[width=1in,height=1.25in,clip,keepaspectratio]{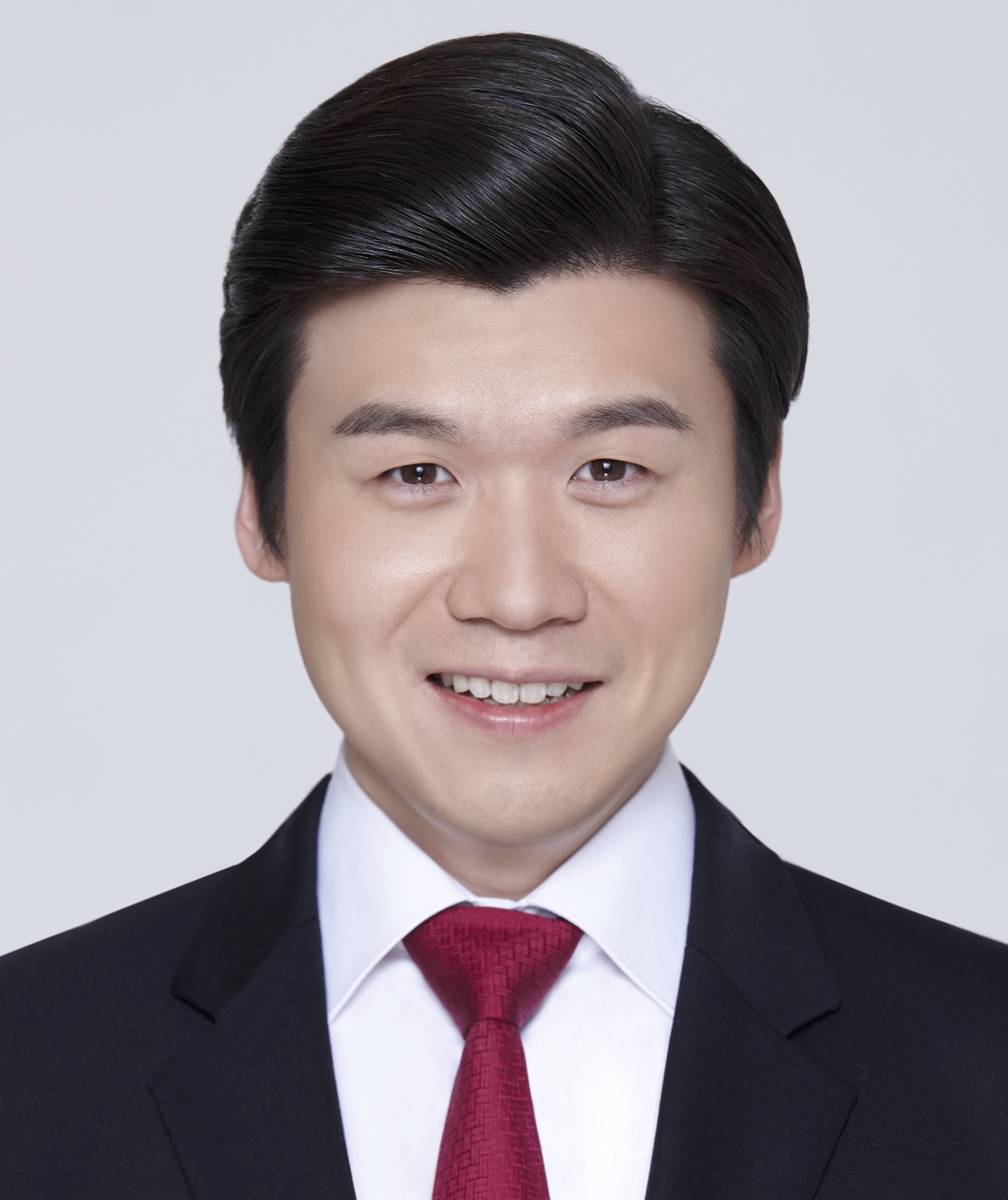}}]{Dr. Insoon Yang} is an Associate Professor of ECE at Seoul National University. He received a Ph.D. in EECS from UC Berkeley in 2015. He was an Assistant Professor of ECE at University of Southern California from 2016 to 2018, and a Postdoctoral Associate with the Laboratory for Information and Decision Systems at Massachusetts Institute of Technology from 2015 to 2016. His research interests are in stochastic control and optimization, and reinforcement learning. He is a recipient of the 2015 Eli Jury Award and a finalist for the Best Student Paper Award at the 55th IEEE Conference on Decision and Control 2016. \end{IEEEbiography}
%
\begin{IEEEbiography}[{\includegraphics[width=1in,height=1.25in,clip,keepaspectratio]{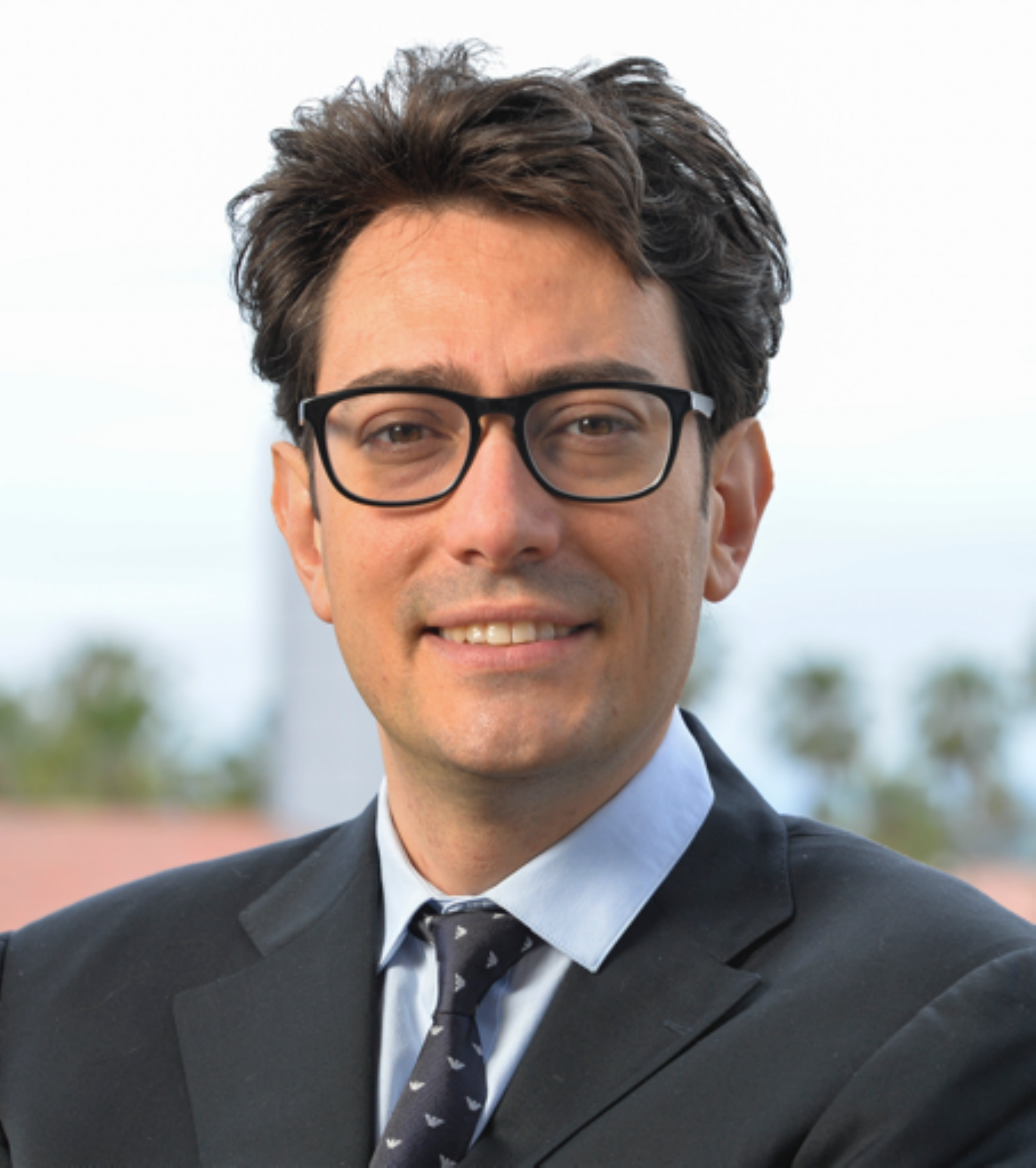}}]{Dr. Marco Pavone} is an Associate Professor of Aeronautics and Astronautics at Stanford University, where he is the Director of the Autonomous Systems Laboratory and Co-Director of the Center for Automotive Research at Stanford. He is currently on a partial leave of absence at NVIDIA serving as Director of Autonomous Vehicle Research. His main research interests are in the development of methodologies for the analysis, design, and control of autonomous systems, with an emphasis on self-driving cars, autonomous aerospace vehicles, and future mobility systems. He is a recipient of a Presidential Early Career Award for Scientists and Engineers (PECASE), an Office of Naval Research Young Investigator Award, a National Science Foundation Early Career (CAREER) Award, a NASA Early Career Faculty Award, and an Early-Career Spotlight
Award from the Robotics Science and Systems Foundation. \end{IEEEbiography}
\vspace{-5mm}
\begin{IEEEbiography}[{\includegraphics[width=1in,height=1.25in,clip,keepaspectratio]{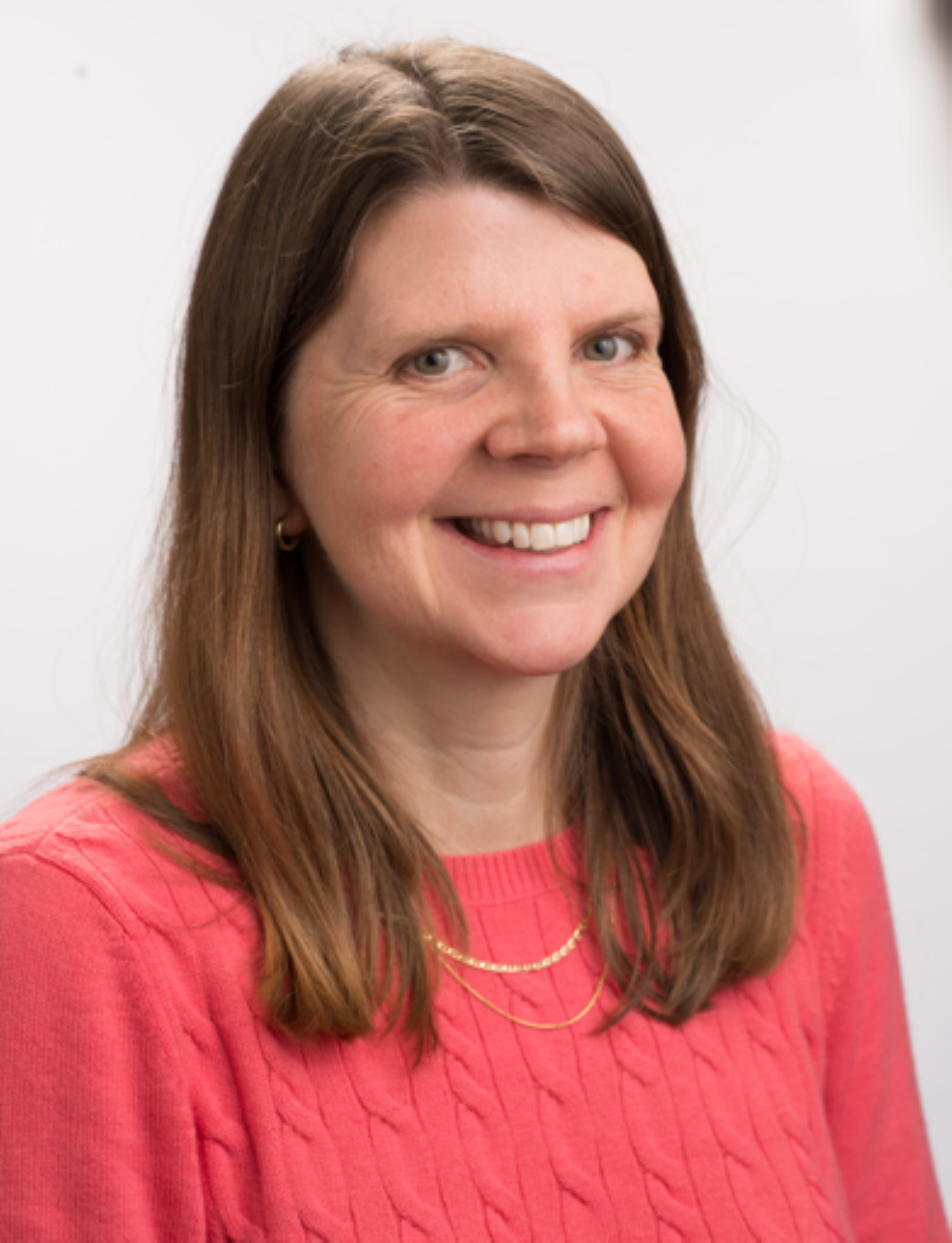}}]{Dr. Claire Tomlin} is the Charles A. Desoer Professor of Engineering in the Department of Electrical Engineering and Computer Sciences (EECS), University of California Berkeley (UC Berkeley). She was an Assistant, Associate, and Full Professor in Aeronautics and Astronautics at Stanford University from 1998 to 2007, and in 2005, she joined UC Berkeley. Claire works in the area of control theory and hybrid systems, with applications to air traffic management, UAV systems, energy, robotics, and systems biology. She is a MacArthur Foundation Fellow (2006), an IEEE Fellow (2010), and in 2017, she was awarded the IEEE Transportation Technologies Award. In 2019, Claire was elected to the National Academy of Engineering and the American Academy of Arts and Sciences. \end{IEEEbiography}

\end{document}


\begin{center}
{\bf Supplementary Material} \\
{\bf ``Risk-sensitive safety analysis using Conditional Value-at-Risk''}\\
{\bf M. Chapman, R. Bonalli, K. Smith, I. Yang, M. Pavone, C. Tomlin} \\[2ex]
\end{center}
\bigskip
\noindent
The purpose of this Supplementary Material is to prove a lemma, whose proof was omitted from the main text due to space constraints. We re-state the following three assumptions from Sec. IV for the reader's convenience:
\begin{assumption}\label{assumption0}
The state space $S$ equals $\mathbb{R}$.
\end{assumption}
\begin{assumption}\label{assumption1}
The control law $\pi_t : \mathbb{R} \rightarrow A$ is continuous.
\end{assumption}
\begin{assumption}\label{assumption2}
The transition law $Q$ satisfies the following convergence property: if the sequence $\{(x^n,u^n)\}_{n \in \mathbb{N}}$ in $\mathbb{R} \times A$ converges to $(x,u) \in \mathbb{R} \times A$, then convergence in total variation holds, $\big|Q(\cdot|x^n,u^n) - Q(\cdot|x,u)\big|(\mathbb{R}) \rightarrow 0$.\footnote{For each $n \in \mathbb{N}$, $Q(\cdot|x^n,u^n) - Q(\cdot|x,u)$ is a signed measure on $(\mathbb{R},\mathcal{B}(\mathbb{R}))$. $|Q(\cdot|x^n,u^n) - Q(\cdot|x,u)|(B)$ denotes the \emph{total variation} of the signed measure evaluated at $B \in \mathcal{B}(\mathbb{R})$.}
\end{assumption}
\noindent
Please see the main paper for justification of these assumptions.
\setcounter{lemma}{4}
\begin{lemma}[Measurability of Multi-Mapping]\label{lemma5}
Let $Y : \mathbb{R} \rightarrow \mathbb{R}$ be a bounded and Borel measurable function. Under Assumptions 1-3, the function\vspace{-2mm}
\begin{equation} \label{eq_multi}
\vspace{-1mm}\begin{aligned}
x_{t-1} \hspace{-1mm} \mapsto \hspace{-1mm}
 \sup_{\xi_t \in \mathcal{R}_{\alpha,t}^\pi(x_{t-1})}  \textstyle\int_{\mathbb{R}} Q(\mathrm{d}x_t |x_{t-1},\pi_{t-1}(x_{t-1})) \xi_t(x_t) Y(x_t)
\end{aligned}\vspace{-1mm}
\vspace{-1mm}\end{equation}
is bounded and Borel measurable.
\end{lemma}
\noindent
The key reason for Assumptions 1-3 is to prove Lemma \ref{lemma5}, which is done below.\\\\
\noindent
\textbf{Proof:} The boundness of \eqref{eq_multi} follows from the definition of $\mathcal{R}_{\alpha,t}^\pi(x_{t-1})$ and from the boundness of $Y$. For its measurability, we first prove that the multi-mapping $\mathcal{G}_t : \mathbb{R} \rightrightarrows \mathbb{R}$\vspace{-2mm}
\begin{equation*}\begin{aligned}
 & x_{t-1}  \mapsto \\
 & \left\{ \hspace{-.5mm} \textstyle \int_{\mathbb{R}}  Q(\mathrm{d}x_t| { x_{t-1},\pi_{t-1}(x_{t-1})}) \xi_t(x_t) Y(x_t) : \xi_t \in \mathcal{R}_{\alpha,t}^\pi(x_{t-1}) \hspace{-.5mm} \right\}
\end{aligned}\vspace{-1mm}\end{equation*}
is measurable. $\mathcal{G}_t : \mathbb{R} \rightrightarrows \mathbb{R}$ means that for any $x_{t-1} \in \mathbb{R}$, $\mathcal{G}_t(x_{t-1}) \subseteq \mathbb{R}$. From \cite[p. 365, Remark 28]{shapiro2009lectures}, to show that $\mathcal{G}_t$ is measurable, it is sufficient to show that it is closed (see \cite[p. 365, Remark 28]{shapiro2009lectures} for this definition of closedness). In particular, we must show that, given a sequence $\{ x^n_{t-1} \}_{n \in \mathbb{N}}$ such that $x^n_{t-1} \longrightarrow x_{t-1}$, $\xi^n_t \in \mathcal{R}^{\pi}_{\alpha,t}(x^n_{t-1})$, and\vspace{-1mm}
\begin{equation*}
\textstyle\int_{\mathbb{R}} Q(\mathrm{d}x_t | x^n_{t-1} , \pi_{t-1}(x^n_{t-1})) \xi^n_t(x_t) Y(x_t) \longrightarrow \ell,
\vspace{-1mm}\end{equation*}
it holds that $\ell \in \mathcal{G}_t(x_{t-1})$. Let $\Tilde{L^2} := L^2\big(\mathbb{R};[0,1 / (\alpha^{\frac{1}{T+1}})];Q(\cdot | { x_{t-1} , \pi_{t-1}(x_{t-1})})\big)$ denote the set of functions in $L^2 := L^2\big(\mathbb{R},\mathcal{B}(\mathbb{R}),Q(\cdot | x_{t-1} , \pi_{t-1}(x_{t-1}))\big)$, such that the co-domain of each function is restricted to $[0,1 / (\alpha^{\frac{1}{T+1}})]$ almost everywhere. Due to the boundness of $\mathcal{R}_{\alpha,t}^\pi(x^n_{t-1})$ for each $n \in \mathbb{N}$, we have $\mathcal{R}_{\alpha,t}^\pi(x^n_{t-1}) \subseteq \tilde{L^2}$ and thus $\{\xi_t^n\}_{n \in \mathbb{N}} \subseteq \tilde{L^2}$. $L^2$ with the $L^2$-norm is a reflexive Banach space, and $\{\xi_t^n\}_{n \in \mathbb{N}}$ is a bounded sequence in $L^2$. Hence, there exists a subsequence, $\{\xi_t^{n_k}\}_{k \in \mathbb{N}}$, that converges to some $\xi_t \in L^2$ in the weak topology of $L^2$ \cite[Theorem 3.18, p. 69]{brezis2010}. Since $\tilde{L^2}$ is \emph{weakly} closed in $L^2$ and $\{\xi_t^{n_k}\}_{k \in \mathbb{N}} \subseteq \tilde{L^2}$ converges weakly to $\xi_t \in L^2$, we have that $\xi_t$ is an element of $\tilde{L^2}$.
\footnote{$L^2$ with the $L^2$-norm is a locally convex topological vector space. Moreover, $\tilde{L^2}$ is a convex subset of $L^2$, and $\tilde{L^2}$ is \emph{strongly} closed in $L^2$. ($\tilde{L^2}$ is \emph{strongly} closed in $L^2$ if and only if for any sequence $\{ f_n : n \in \mathbb{N}\}$ in $\tilde{L^2}$ such that $f_n$ converges strongly to $f \in L^2$, we have $f \in \tilde{L^2}$. $f_n$ converges \emph{strongly} to $f$ if and only if the $L^2$-norm of $f_n - f$ tends to 0 as $n$ tends to infinity.) Thus, $\tilde{L^2}$ is \emph{weakly} closed in $L^2$ \cite[Theorem 3.5.15, p. 162]{ash1972}. ($\tilde{L^2}$ is \emph{weakly} closed in $L^2$ if and only if for any sequence $\{ f_n : n \in \mathbb{N}\}$ in $\tilde{L^2}$ such that $f_n$ converges weakly to $f \in L^2$, we have $f \in \tilde{L^2}$. $f_n$ converges \emph{weakly} to $f$ if and only if $\sup\{ ||f_n||_2 : n \in \mathbb{N} \}$ is finite, where $||\cdot||_2$ denotes the $L^2$-norm, and $\int_B f_n \mathrm{d}Q(\cdot|x_{t-1},\pi_{t-1}(x_{t-1}))$ tends to $\int_B f \mathrm{d}Q(\cdot|x_{t-1},\pi_{t-1}(x_{t-1}))$ as $n$ tends to infinity for any $B \in \mathcal{B}(\mathbb{R})$ \cite[Theorem 3.4.13, p. 160]{ashprob2000}.) Recall that in our setting $L^2 := L^2\big(\mathbb{R},\mathcal{B}(\mathbb{R}),Q(\cdot | x_{t-1} , \pi_{t-1}(x_{t-1}))\big)$.} Next, we show that $\xi_t \in \mathcal{R}^{\pi}_{\alpha,t}(x_{t-1})$. Denote $Q_{\pi}(\cdot|y) := Q(\cdot| y , \pi_{t-1}(y))$ for any $y \in \mathbb{R}$. Then,
\begin{align*}\vspace{-1mm}
    & \textstyle \left| {   \int_\mathbb{R} Q_\pi(\mathrm{d}x_t | x_{t-1}) \xi_t(x_t) - 1 } \right| \\
    &\le  \textstyle \left| {  \int_\mathbb{R} Q_\pi(\mathrm{d}x_t | x_{t-1}) \big( \xi_t(x_t) - \xi^{n_k}_t(x_t) \big) } \right| \\
    & \textstyle \textcolor{white}{\le} + \left| { \int_\mathbb{R} Q_\pi(\mathrm{d}x_t | x_{t-1}) \xi^{n_k}_t(x_t) - 1  } \right|\\
    & \textstyle \leq \left| { \int_\mathbb{R} Q_\pi(\mathrm{d}x_t | x_{t-1}) \big( \xi_t(x_t) - \xi^{n_k}_t(x_t) \big) } \right|  \\
    &\textstyle \quad + | \int_\mathbb{R} Q_\pi(\mathrm{d}x_t | x_{t-1}) \xi^{n_k}_t(x_t) - \textstyle \int_\mathbb{R} Q_\pi(\mathrm{d}x_t | x^{n_k}_{t-1}) \xi^{n_k}_t(x_t) | \\
    &\le \textstyle \left| \int_\mathbb{R} Q_\pi(\mathrm{d}x_t | x_{t-1}) \big( \xi_t(x_t) - \xi^{n_k}_t(x_t) \big) \right| \\
    &\quad + \textstyle\frac{1}{\alpha^{\frac{1}{T+1}}} \big| Q_\pi( \cdot |  x_{t-1} ) - Q_\pi( \cdot |   x^{n_k}_{t-1} ) \big|(\mathbb{R}),
\vspace{-1mm}\end{align*}
which tends to zero as $k \longrightarrow \infty$, since $\xi_t^{n_k}$ converges \emph{weakly} to $\xi_t$ in $L^2$ and by Assumptions \ref{assumption1} and \ref{assumption2}.\footnote{Since $\pi_{t-1}$ is continuous by Assumption \ref{assumption1} and $x_{t-1}^{n_k} \rightarrow x_{t-1}$ as $k \rightarrow \infty$, then $(x_{t-1}^{n_k}, \pi_{t-1}(x_{t-1}^{n_k})) \rightarrow (x_{t-1}, \pi_{t-1}(x_{t-1}))$ as $k \rightarrow \infty$. Thus, by Assumption \ref{assumption2}, the transition laws converge in total variation.} Since $\int_\mathbb{R} Q(\mathrm{d}x_t | x_{t-1} , \pi_{t-1}(x_{t-1})) \xi_t(x_t) = 1$ and $\xi_t \in \tilde{L^2}$, we have that $\xi_t \in \mathcal{R}^{\pi}_{\alpha,t}(x_{t-1})$. By leveraging similar computations and the boundness of $Y$, we have\vspace{-1mm}
\begin{align*}
    \textstyle \int_{\mathbb{R}} \hspace{-0.5mm}Q_\pi(\mathrm{d}x_t |  x^{n_k}_{t-1}  ) \xi^{n_k}_t(x_t) Y(x_t) 
    \rightarrow \int_{\mathbb{R}} \hspace{-0.5mm} Q_\pi(\mathrm{d}x_t |  x_{t-1} ) \xi_t(x_t) Y(x_t),
\vspace{-1mm}\end{align*}
as $k \longrightarrow \infty$, which gives\vspace{-1mm}
\begin{equation*}
\textstyle \ell = \int_{\mathbb{R}} Q(\mathrm{d}x_t | x_{t-1} , \pi_{t-1}(x_{t-1})) \xi_t(x_t) Y(x_t) \in \mathcal{G}_t(x_{t-1}).
\vspace{-1mm}\end{equation*}
We are using the fact that if a sequence in $\mathbb{R}$ converges, then any subsequence converges to the limit of the original sequence. Since $\ell \in \mathcal{G}_t(x_{t-1})$, the multi-mapping $\mathcal{G}_t$ is closed, which implies that $\mathcal{G}_t$ is measurable and closed-valued \cite[p. 365, Remark 28]{shapiro2009lectures}. The domain of $\mathcal{G}_t$, $\text{dom }\mathcal{G}_t := \{ x_{t-1} \in \mathbb{R} : \mathcal{G}_t(x_{t-1}) \text{ is not empty} \}$, equals $\mathbb{R}$. Indeed, for any $x_{t-1} \in \mathbb{R}$, $\mathcal{G}_t(x_{t-1})$ is not empty because the function equal to 1 everywhere is an element of $\mathcal{R}^{\pi}_{\alpha,t}(x_{t-1})$, so $\int_{\mathbb{R}} Q(\mathrm{d}x_t | x_{t-1} , \pi_{t-1}(x_{t-1})) Y(x_t) \in \mathcal{G}_t(x_{t-1})$. In summary, the multi-mapping $\mathcal{G}_t : \mathbb{R} \rightrightarrows \mathbb{R}$ is closed-valued and measurable with $\text{dom }\mathcal{G}_t = \mathbb{R}$. Thus, by \cite[p. 646, Theorem 14.5 (a)]{rockwets}, there exists a countable family of Borel measurable functions $g_n : \mathbb{R} \rightarrow \mathbb{R}$ such that, for every $x_{t-1} \in \mathbb{R}$, we have $\mathcal{G}_t(x_{t-1}) = \textnormal{cl}  \{ g_n(x_{t-1}) \}_{n \in \mathbb{N}}$, where cl denotes the closure.\footnote{\textcolor{blue}{Here, we provide the statement of \cite[p. 646, Theorem 14.5 (a)]{rockwets} (using some different symbols to avoid confusion with other notation in this paper). Let $(\bar{\Omega}, \bar{\mathcal{F}})$ be a measurable space. The measurability of a closed-valued mapping $\bar{\mathcal{G}} : \bar{\Omega} \rightrightarrows \mathbb{R}^m$ is equivalent to (in combination with the measurability of $\text{dom }\bar{\mathcal{G}}$): there is a countable family $\{ g_j\}_{j \in \mathbb{N}}$ of measurable functions $g_j : \text{dom }\bar{\mathcal{G}} \rightarrow \mathbb{R}^m$ such that $\bar{\mathcal{G}}(\bar{\omega}) = \textnormal{cl}\{ g_j(\bar{\omega}) : j \in \mathbb{N}\}$ for each $\bar{\omega} \in \bar{\Omega}$.}} Hence, for every $x_{t-1} \in \mathbb{R}$,\\
    $\sup \mathcal{G}_t(x_{t-1}) = \sup \ \textnormal{cl} \{ g_n(x_{t-1}) \}_{n \in \mathbb{N}} = \underset{n \in \mathbb{N}}{\sup} \ g_n(x_{t-1}).$
Thus, the function $\sup \mathcal{G}_t : \mathbb{R} \longrightarrow \mathbb{R}$ is Borel measurable. $\blacksquare$